\documentclass[aps,prd,superscriptaddress,showpacs,preprintnumbers,floatfix]{revtex4}
\usepackage{graphicx,color}
\newcommand{\be}{\begin{equation}}
\newcommand{\ee}{\end{equation}}
\newcommand{\bea}{\begin{eqnarray}}
\newcommand{\eea}{\end{eqnarray}}

\newcommand{\bfk}{\mbox{\boldmath $k$}}
\newcommand{\bfb}{\mbox{\boldmath $b$}}

\newcommand{\bfp}{\mbox{\boldmath $p$}}

\newcommand{\bfS}{\mbox{\boldmath $S$}}

\newcommand{\pup}{p^\uparrow}

\newcommand{\avkt}{\langle k_\perp^2 \rangle}
\newcommand{\avpt}{\langle p_\perp^2 \rangle}
\def\lsim{\mathrel{\rlap{\lower4pt\hbox{\hskip1pt$\sim$}}\raise1pt\hbox{$<$}}}
\def\gsim{\mathrel{\rlap{\lower4pt\hbox{\hskip1pt$\sim$}}\raise1pt\hbox{$>$}}}
\def\nostrocostruttino#1\over#2{\mathrel{\mathop{\kern 0pt \rlap
{\hbox{$#1$}}} \hbox{\kern-.135em $#2$}}}

\def\kt{k_\perp}
\def\bt{b_T}
\def\bkt{\bfk_\perp}
\def\bbt{\bfb_T}

\def\pp{p_\perp}
\def\bpp{\bfp_\perp}

\def\avp{\langle p_\perp ^2\rangle}

\begin{document}
%
\title{A strategy towards the extraction of the Sivers function 
with TMD evolution}
\author{M.~Anselmino}
\affiliation{Dipartimento di Fisica Teorica, Universit\`a di Torino,
             Via P.~Giuria 1, I-10125 Torino, Italy}
\affiliation{INFN, Sezione di Torino, Via P.~Giuria 1, I-10125 Torino, Italy}
\author{M.~Boglione}
\affiliation{Dipartimento di Fisica Teorica, Universit\`a di Torino,
             Via P.~Giuria 1, I-10125 Torino, Italy}
\affiliation{INFN, Sezione di Torino, Via P.~Giuria 1, I-10125 Torino, Italy}
%
%
\author{S.~Melis}
\affiliation{European Centre for Theoretical Studies in Nuclear Physics 
             and Related Areas (ECT*), \\
             Villa Tambosi, Strada delle Tabarelle 286, I-38123 Villazzano, 
             Trento, Italy}
%
%

\begin{abstract}
The QCD evolution of the unpolarized Transverse Momentum Dependent (TMD) 
distribution functions and of the Sivers functions have been discussed in 
recent papers. 
Following such results we reconsider previous extractions of the Sivers 
functions from semi-inclusive deep inelastic scattering data and propose 
a simple strategy which allows to take into account the $Q^2$ dependence 
of the TMDs in comparison with experimental findings. A clear evidence of 
the phenomenological success of the TMD evolution equations is given, mostly, 
by the newest COMPASS data off a transversely polarized proton target.        
\noindent
\end{abstract}

\pacs{13.88.+e, 13.60.-r, 13.85.Ni}

\maketitle

\section{\label{Intro} Introduction and formalism}

The exploration of the 3-dimensional structure of the nucleons, both 
in momentum and configuration space, is one of the major issues in hadron 
high energy physics, with dedicated experimental and theoretical efforts. 
In particular, several Semi-Inclusive Deep Inelastic Scattering (SIDIS) 
experiments are either running or being planned. From the measurements of 
azimuthal asymmetries, both with unpolarized and polarized nucleons, one 
obtains information on the Transverse Momentum Dependent Parton Distribution 
Functions (TMD PDFs) and on the Transverse Momentum Dependent Fragmentation 
Functions (TMD FFs). The TMD PDFs and the TMD FFs are often globally 
referred to simply as TMDs. The TMD PDFs convey information on the momentum 
distributions of partons inside protons and neutrons.   

The analysis of the experimental data is based on the so-called TMD 
factorization, which links measurable cross sections and spin asymmetries 
to a convolution of TMDs. In particular, the Sivers function, which describes
the number density of unpolarized quarks inside a transversely polarized 
proton, has received much attention and has been extracted from SIDIS data
by several groups, with consistent results \cite{Anselmino:2005ea,
Collins:2005ie,Vogelsang:2005cs,Anselmino:2005an,Anselmino:2008sga,
Bacchetta:2011gx}. However, all these phenomenological fits of the Sivers 
function (and other TMDs) have been performed so far using a simplified 
version of the TMD factorization scheme, in which the QCD scale dependence 
of the TMDs -- which was unknown -- is either neglected or limited to the 
collinear part of the unpolarized PDFs. While this might not be a serious 
numerical problem when considering only experimental data which cover 
limited ranges of low $Q^2$ values, it is not correct in principle, and 
taking into account the appropriate $Q^2$ evolution might be numerically 
relevant for predictions at higher $Q^2$ values, like future electron-ion 
or electron-nucleon colliders (EIC/ENC) and Drell-Yan experiments.            

Recently, the issue of the QCD evolution of unpolarized TMDs and of 
the Sivers function has been studied in a series of papers 
\cite{Collins:2011book, Aybat:2011zv,Aybat:2011ge} and a 
complete TMD factorization framework is now available for a consistent 
treatment of SIDIS data and the extraction of TMDs. A first application
of the new TMD evolution equations to some limited samples of the HERMES 
and COMPASS data~\cite{Aybat:2011ta} has indeed shown clear signs of  
the $Q^2$ TMD evolution.     
  
We follow here Refs. \cite{Aybat:2011zv} and \cite{Aybat:2011ge} adopting
their formalism, which includes the explicit $Q^2$ dependence of the TMDs, 
and apply it to the extraction of the Sivers function from SIDIS data, 
exploiting the latest HERMES~\cite{:2009ti} and 
COMPASS~\cite{Bradamante:2011xu} results. 
In the sequel of this Section we present the explicit formalism: in 
Subsection~\ref{Intro-sub} we describe the setup and structure of the 
TMD evolution equations, in Subsection~\ref{Param} we discuss the
parameterizations used for the unknown input functions, while in 
Subsection~\ref{gaussian} we present analytical solutions of the TMD 
evolutions equations obtained under a specific approximation.

In Section~\ref{newfit} we perform a best fit of the SIDIS Sivers 
asymmetries taking into account the different $Q^2$ values of each data 
point and the $Q^2$ dependence of the TMDs; we compare our results with
a similar analysis performed without the TMD evolution. Differences between 
Sivers functions extracted from data with and without the TMD evolution 
are shown and commented. In all this we differ from Ref.~\cite{Aybat:2011ta}, 
which explicitly shows the evolution of an existing fit of the Sivers SIDIS 
asymmetry \cite{Anselmino:2011gs} from the average value $\langle Q^2 
\rangle = 2.4$~GeV$^2$ for HERMES data~\cite{:2009ti} to the average 
value of $\langle Q^2 \rangle = 3.8$ GeV$^2$ for the most recent 
COMPASS data~\cite{Bradamante:2011xu}. 
Further comments and conclusions are given in Section~\ref{conc}.    
   
\subsection{\label{Intro-sub} Formalism for TMD $Q^2$ dependence}    

In Refs.~\cite{Collins:2011book} and~\cite{Aybat:2011zv}, Collins, 
Aybat and Rogers have proposed a scheme to describe the $Q^2$ evolution 
of the TMD unpolarized distribution and fragmentation functions: within 
the framework of the Collins-Soper-Sterman (CSS) factorization 
formalism~\cite{Collins:1981va,Collins:1984kg}, they can describe the 
non-perturbative, low transverse momentum region and, at the same time, 
consistently include the perturbative corrections affecting the region 
of larger energies and momentum transfers. However, this formalism cannot 
be directly applied to spin dependent distribution functions, like the 
Sivers function~\cite{Sivers:1989cc}, for which the collinear 
limit does not exist. 

More recently, an extension of the unpolarized TMD-evolution formalism was presented in Ref.~\cite{Aybat:2011ge} to provide a framework in which 
also spin-correlated PDFs can be accounted for. For our purposes, we will 
use Eq.~(44) of Ref.~\cite{Aybat:2011ge} which, compared to the unpolarized 
TMD evolution scheme, Eq.~(26) of Ref.~\cite{Aybat:2011zv}, requires the 
extra aid of a phenomenological input function embedding the missing 
information on the evolved function, that, in the case of the Sivers function, 
is both of perturbative and non-pertubative nature. Although the unpolarized 
PDF and FF TMD evolution equations are in principle known~\cite{Aybat:2011zv}, 
in this paper we adopt the simplified functional form of the evolution 
equation, as proposed for the Sivers function in Ref.~\cite{Aybat:2011ge}, 
for all TMD functions, for consistency. 

Thus, we strictly follow Ref.~\cite{Aybat:2011ge} and combine their Eqs.~(44), 
(43) and (30), taking, as suggested~\cite{Aybat:2011ge}, the renormalization 
scale $\mu^2$ and the regulating parameters $\zeta_F$ and $\zeta_D$ all equal 
to $Q^2$. Then, the QCD evolution of the TMDs in the coordinate space can be written as  
\be
\widetilde F(x, \bbt; Q) = \widetilde F(x, \bbt; Q_0)\exp \left\{
\ln \frac{Q}{Q_0} \> \widetilde K(b_T; Q_0)
+ \int_{Q_0}^Q \frac{\rm d \mu}{\mu} \gamma_F \left( \mu, \frac{Q^2}{\mu^2}
\right) \right\} \>, \label{evol}
\ee
where $\widetilde F$ can be either the unpolarized parton distribution,
$\widetilde F(x, \bbt; Q) = \widetilde f_{q/p}(x, \bbt; Q)$, the unpolarized 
fragmentation function $\widetilde F(x, \bbt; Q) = \widetilde D_{h/q}(z, 
\bbt; Q)$, or the first derivative, with respect to the parton impact 
parameter $b_T$, of the Sivers function, $\widetilde F(x, \bbt; Q) = 
\widetilde f_{1T}^{\prime \perp f}(x, \bbt; Q)$. Notice that throughout
the paper $\bt$-dependent distribution and fragmentation functions will be 
denoted with a $\sim$ on top.

In the above equation the function $\widetilde K$ is given in general 
by \cite{Aybat:2011ge}:  
\be
\widetilde K(\bt, \mu) = \widetilde K(b_*, \mu_b) + \left[\int_{\mu}^{\mu_b}
\frac{\rm d \mu'}{\mu'} \gamma_K(\mu') \right]- g_K(b_T) \>, \label{Kt}
\ee  
with, at ${\cal O}(\alpha_s)$ \cite{Collins:1981va,Collins:1984kg}, 
\be
\widetilde K(b_*, \mu_b) = - \frac{\alpha_s \, C_F}{\pi} \left[
\ln(b_*^2 \, \mu_b^2) - \ln 4 + 2 \gamma_E \right]
\ee
\be
b_*(b_T) \equiv \frac{b_T}{\sqrt{1 + b_T^2/b_{\rm max}^2}} 
\quad\quad\quad
\mu_b = \frac{C_1}{b_*(b_T)} \> \cdot \label{mub}
\ee
  
The first two terms in Eq.~(\ref {Kt}) are perturbative and depend on 
the scale $\mu$ through the coupling $\alpha_s(\mu)$, while the last 
term is non-perturbative, but scale independent. $C_1$ is a constant 
parameter which can be fixed to optimize the perturbative expansion, 
as explained in Ref.~\cite{Collins:1984kg}. Refs.~\cite{Aybat:2011zv} 
and~\cite{Aybat:2011ge} adopt the particular choice $C_1 = 2 e^{-\gamma_E}$ 
which automatically implies $\widetilde K(b_*, \mu_b) = 0$, considerably simplifying the $\bt$ dependence of the CSS kernel 
$\widetilde K(\bt, \mu)$, Eq.~(\ref{Kt}).

The anomalous dimensions $\gamma_F$ and $\gamma_K$ appearing respectively 
in Eqs. (\ref{evol}) and (\ref{Kt}), are given, again at order 
${\cal O}(\alpha_s)$, by~\cite{Collins:1984kg,Aybat:2011zv}
\be
\gamma_F(\mu; \frac{Q^2}{\mu^2}) = \alpha_s(\mu) \, \frac{C_F}{\pi}
\left( \frac{3}{2} - \ln \frac{Q^2}{\mu^2} \right)
\quad\quad\quad\quad 
\gamma_K(\mu) = \alpha_s(\mu) \, \frac{2 \, C_F}{\pi} \> \cdot
\label{gammas}
\ee
    
By making use of Eqs.~(\ref{Kt})-(\ref{gammas}), the evolution  of 
$\widetilde F(x, \bbt; Q)$ in Eq.~(\ref{evol}) can then be written as:   
\be
\widetilde F(x, \bbt; Q) = \widetilde F(x, \bbt; Q_0)\> 
\widetilde R(Q, Q_0, \bt)\> \exp \left\{- g_K(b_T) \ln \frac{Q}{Q_0} \right\} 
\>, \label{Ftev}
\ee
with
\be
\widetilde R(Q, Q_0, \bt)
\equiv 
\exp \left\{ \ln \frac{Q}{Q_0} \int_{Q_0}^{\mu_b} \frac{\rm d \mu'}{\mu'} \gamma_K(\mu') +
\int_{Q_0}^Q \frac{\rm d \mu}{\mu} 
\gamma_F \left( \mu, \frac{Q^2}{\mu^2} \right)\right\} 
\> \cdot \label{RQQ0}
\ee
The $Q^2$ evolution is driven by the functions $g_K(b_T)$ and 
$\widetilde R(Q,Q_0,\bt)$. While the latter, Eq.~(\ref{RQQ0}), can be easily evaluated, numerically or even analytically, the former, is essentially 
unknown and will need to be taken from independent experimental inputs.

The explicit expression of the TMDs in the momentum space, with the QCD 
$Q^2$ dependence, can be obtained by Fourier-transforming Eq.~(\ref{Ftev}),
obtaining~\cite{Aybat:2011ge}: 
\be
\widehat f_{q/p}(x, \kt; Q) = \frac{1}{2\pi} \int_0^\infty \!\!\!{\rm d} b_T 
\> b_T \> J_0(k_\perp b_T) \> \widetilde f_{q/p}(x, b_T; Q) 
\label{TMDunpf}
\ee
\be
\widehat D_{h/q}(z, \pp; Q) = \frac{1}{2\pi} \int_0^\infty \!\!\!{\rm d} b_T 
\> b_T \> J_0({\rm k}_T b_T) \> \widetilde D_{h/q}(z, \bt; Q) 
\label{TMDunpD}
\ee
\be
\widehat f_{1T}^{\perp f}(x, k_\perp; Q) = \frac{-1}{2\pi k_\perp} \int_0^\infty 
\!\!\! {\rm d} b_T \> b_T \> J_1(k_\perp b_T) \> 
\widetilde f_{1T}^{\prime \,\perp q}(x, b_T; Q) \>, \label{TMDsiv}
\ee
where $J_0$ and $J_1$ are Bessel functions. In this paper we denote the distribution and fragmentation functions which depend on the transverse 
momenta (TMDs) with a ``widehat" on top. $\widehat f_{q/p}$ is the unpolarized 
TMD distribution function for a parton of flavor $q$ inside a proton, and 
$\widehat D_{h/q}$ is the unpolarized TMD fragmentation function for hadron $h$ 
inside a parton $q$. $\widehat f_{1T}^{\perp q}$ is the Sivers distribution 
defined, for unpolarized partons inside a transversely polarized proton, as:
\bea
\widehat f_{q/p^\uparrow}(x, \bfk_\perp, \bfS; Q) &=& 
\widehat f_{q/p}(x, k_\perp; Q) - \widehat f_{1T}^{\perp q}(x, k_\perp; Q)\frac{\epsilon_{ij} \, k_\perp^i \, 
S^j}{M_p} \label{Siv1} \\
&=& \widehat f_{q/p}(x, k_\perp; Q)
+ \frac 12 \Delta^N \widehat f_{q/p^\uparrow}(x, k_\perp; Q)\frac{\epsilon_{ij} 
\, k_\perp^i \, S^j}{k_\perp} \> \cdot \label{Siv2}
\eea

In our notation $\bkt$ is the transverse momentum of the parton with 
respect to the parent nucleon direction and $\bpp$ is the transverse 
momentum of the final hadron with respect to the parent parton direction.   
Notice that in Refs.~\cite{Aybat:2011zv} and~\cite{Aybat:2011ge} all 
transverse momenta are defined in a unique frame, the so-called hadron 
frame, in which the measured hadrons have zero transverse momentum. 
In this frame, the initial and the final parton transverse momenta are 
denoted, respectively, by ${\bf k}_{1T}$ and ${\bf k}_{2T}$. They are related 
to our notation by: $\bkt = {\bf k}_{1T}$ and, at leading order in $p_\perp$, 
$\bpp = - z \, {\bf k}_{2T}$. This requires some attention when dealing with 
the fragmentation functions. Usually, the TMD FFs are defined in terms of the
{\it hadronic} $\pp$, i.e. the transverse momentum of the final hadron $h$ 
with respect to the direction of the fragmenting parton $q$, while, 
following Refs.~\cite{Aybat:2011zv} and~\cite{Aybat:2011ge}, the Fourier 
transform (\ref{TMDunpD}) is performed from the impact parameter space of the fragmenting parton ($b_T$) into the corresponding {\it partonic} transverse momentum (k$_T = \pp / z$) in the hadron frame. This will generate some extra 
$z^2$ factors, as explained in detail in Section~\ref{Param}. 

\subsection{\label{Param} Parameterization of unknown functions}    

Eqs.~(\ref{TMDunpf})-(\ref{TMDsiv}) can be adopted as the appropriate 
functional forms, with the correct $Q^2$ dependence induced by 
Eqs.~(\ref{Ftev})-(\ref{RQQ0}), to be used in the extraction of 
phenomenological information on the unpolarized and Sivers TMDs.
In order to do so, one should start with a parameterization of the unknown 
functions inside Eq.~(\ref{Ftev}): $g_K(\bt)$ and $\widetilde F(x, b_T; Q_0)$.
As already anticipated, $g_K(\bt)$ is a non-perturbative, but universal function, which in the literature is usually parameterized in a quadratic form. As in 
Refs.~\cite{Aybat:2011ge} and \cite{Aybat:2011ta}, we will adopt the results provided by a recent fit of Drell-Yan data~\cite{Landry:2002ix}, and assume
\be
g_K(b_T) = \frac12 \, g_2 \, b_T^2 \quad\quad {\rm with} \quad\quad 
g_2 = 0.68  \quad\quad 
{\rm corresponding~to} \quad\quad b_{\rm max}=0.5 \> {\rm GeV}^{-1} \>.
\label{gk}
\ee

We should now parameterize the function $\widetilde F(x, b_T; Q_0)$ in 
configuration space. We wish to test the effect of the TMD evolution
in the extraction of the Sivers functions from data; in particular we will  
compare the extraction based on TMD evolution with previous extractions 
which did not take such an evolution into account. Then, we parameterize the 
input function $\widetilde F(x, b_T; Q_0)$ by requiring that its 
Fourier-transform, which gives the corresponding TMD function in the 
transverse momentum space, coincides with the previously adopted 
$\kt$-Gaussian form, with the $x$ dependence factorized out. That was also 
done in Refs.~\cite{Aybat:2011zv} and~\cite{Aybat:2011ge}, assuming for the 
unpolarized TMD PDF
\be
\widetilde f_{q/p}(x, \bt; Q_0) = f_{q/p}(x,Q_0) \exp \left\{-\alpha ^2\,b_T^2   \right\} \>, \label{Fbt}
\ee  
where $f_{q/p}(x,Q_0)$ is the usual integrated PDF of parton $q$ inside 
proton $p$, evaluated at $Q_0$; the value of $\alpha ^2$ is fixed by 
requiring the desired behavior of the distribution function in the transverse momentum space at the initial scale $Q_0$: taking $\alpha ^2=\avkt/4$ one 
recovers
\be
\widehat f_{q/p}(x,k_\perp; Q_0) = f_{q/p}(x,Q_0) \, \frac{1}{\pi 
\langle\kt^2\rangle} \,
e^{-{\kt^2}/{\langle\kt^2\rangle}}\,,\label{partonf}
\ee
in agreement with Refs. \cite{Anselmino:2008sga,Anselmino:2011gs,
Anselmino:2011ch}. 

Similar relations hold for the TMD FFs, with an additional $z^2$ factor due 
to the fact that the Fourier-transform~(\ref{TMDunpD}) leads from the impact 
parameter space of the fragmenting parton in the hadron frame to the 
corresponding partonic transverse momentum k$_T$, while the TMD FFs are 
functions of the transverse momentum $\pp = z \, {\rm k}_T$ of the 
final hadron with respect to the fragmenting parton direction. This requires 
the initial parameterization     
\be
\widetilde D_{h/q}(z, \bt; Q_0) = \frac{1}{z^2} \, D_{h/q}(z,Q_0) \; 
\exp \left\{-\beta ^2\,b_T^2   \right\} \>, \label{Dbt}
\ee 
where $D_{h/q}(z,Q_0)$ is the usual integrated FF evaluated at the initial 
scale $Q_0$, and $\beta ^2 = \avp/4z^2$ in order to recover the previously 
adopted behavior~\cite{Anselmino:2008sga,Anselmino:2011gs,Anselmino:2011ch} 
of the fragmentation function in the $\pp$ transverse momentum space at $Q_0$:
\be
\widehat D_{h/q}(z,p _\perp; Q_0) = D_{h/q}(z,Q_0) \, \frac{1}
{\pi \langle p_\perp^2\rangle}
\, e^{-p_\perp^2/\langle p_\perp^2\rangle} \>.
\label{partond}
\ee
Analogously, we parameterize the Sivers function at the initial scale $Q_0$ as
\be
\label{sivers'}
\widetilde f_{1T}^{\prime \perp}(x, \bt; Q_0 ) = - 2 \, \gamma^2 \, 
f_{1T}^{\perp}(x; Q_0 ) \, \bt \, e^{-\gamma^2 \, b_T^2} \label{Sbt} \>,
\ee
which, when Fourier-transformed according to Eq.~(\ref{TMDsiv}), yields: 
\be
\widehat f_{1T}^{\perp}(x, \kt; Q_0) = f_{1T}^{\perp }(x; Q_0 ) \, \frac{1}
{4 \, \pi \, \gamma^2} \, e^{-\kt^2 / 4 \gamma^2} \>. \label{sivers}
\ee
Eq.~(\ref{sivers}) agrees with our previous parameterization of the Sivers 
function, at the initial scale $Q_0$~\cite{Anselmino:2008sga,Anselmino:2011gs,Anselmino:2011ch}, taking:
\be
4 \, \gamma^2 \equiv \avkt _S = \frac{M_1^2 \, \avkt}{M_1^2 +\avkt}
\label{gamma}
\ee
\vskip -12pt
\be
f_{1T}^{\perp}(x; Q_0) = - \frac {M_p}{2 M_1} \sqrt{2e} \;
\Delta^N \! f_{q/p^\uparrow}(x,Q_0) \, \frac{\avkt_S}{\avkt} \>\cdot
\label{fiTp}
\ee
$M_1$ is a mass parameter, $M_p$ the proton mass and $\Delta^N \! 
f_{q/p^\uparrow}(x,Q_0)$ is the $x$-dependent term of the Sivers function, evaluated at the initial scale $Q_0$ and written as~\cite{Anselmino:2008sga,Anselmino:2011gs,Anselmino:2011ch}:
\be
\Delta^N \! f_{q/p^\uparrow}(x,Q_0) =2 \, {\cal N}_q(x) \, f_{q/p} (x,Q_0)
\; , \label{DeltaN}
\ee
where ${\cal N}_q(x)$ is a function of $x$, properly parameterized  
(we will come back to details of the Sivers function parameterization in 
Section~\ref{newfit}). 
 
The final evolution equations of the unpolarized TMD PDFs and TMD FFs, in
the configuration space, are obtained inserting Eqs.~(\ref{Fbt}) and
(\ref{Dbt}) into Eq.~(\ref{Ftev}):  
\be
\widetilde f_{q/p}(x, \bt; Q) =  f_{q/p}(x,Q_0) \;\widetilde R(Q, Q_0, \bt)\;
\exp \left\{-b_T^2 \left(\alpha ^2\,  + \frac{g_2}{2} \ln \frac{Q}{Q_0}\right) \right\} 
\label{evF-f}
\ee
\vskip -12pt
\be
\widetilde D_{h/q}(z, \bt; Q) = \frac{1}{z^2} D_{h/q}(z,Q_0) \;
\widetilde R(Q, Q_0, \bt)\;
\exp \left\{- b_T^2 \left(\beta ^2\,  + \frac{g_2}{2} 
\ln \frac{Q}{Q_0}\right) \right\} \>,
\label{evF-D}
\ee
with $\alpha^2 = \avkt/4$, $\beta^2 = \avpt/(4z^2)$, $g_2$ given in 
Eq.~(\ref{gk}) and $\widetilde R(Q, Q_0, \bt)$ in Eq.~(\ref{RQQ0}).

The evolution of the Sivers function is obtained through its first 
derivative, inserting Eq.~(\ref{Sbt}) into Eq.~(\ref{Ftev}):  
\be
\widetilde f_{1T}^{\prime \perp }(x, \bt; Q) =  -2 \, \gamma^2 \,
f_{1T}^{\perp}(x; Q_0) \, \widetilde R(Q,Q_0,b_T) \, \bt \, 
\exp \left\{-b_T^2 \left( \gamma^2\, + \frac{g_2}{2} \ln \frac{Q}{Q_0} 
\right) \right\} \,
\label{evF-Sivers}
\ee
with $\gamma^2$ and $f_{1T}^{\perp}(x; Q_0)$ given in 
Eqs.~(\ref{gamma})-(\ref{DeltaN}).

Eqs.~(\ref{evF-f})-(\ref{evF-Sivers}) show that the $Q^2$ evolution is 
controlled by the logarithmic $Q$ dependence of the $b_T$ Gaussian width, 
together with the factor $\widetilde R(Q, Q_0, \bt)$: for increasing values 
of $Q^2$, they are responsible for the typical broadening effect already 
observed in Refs.~\cite{Aybat:2011zv} and~\cite{Aybat:2011ge}.

It is important to stress that although the structure of Eq.~(\ref{evol}) 
is general and holds over the whole range of $\bt$ values, the input function 
$\widetilde F(x,\bbt,Q_0)$ is only designed to work in the large-$b_T$ 
region, corresponding to low $\kt$ values. Therefore, this formalism is 
perfectly suitable for phenomenological applications in the kinematical 
region we are interested in, but the parameterization of the input function 
should be revised in the case one wishes to apply it to a wider range of
transverse momenta, like higher $Q^2$ processes  where perturbative 
corrections become important.   

\subsection{\label{gaussian} Analytical solution of the TMD Evolution Equations}
  
\begin{figure}[t]
\begin{center}\hspace*{-1cm}
\includegraphics[width=0.3\textwidth,angle=-90]
{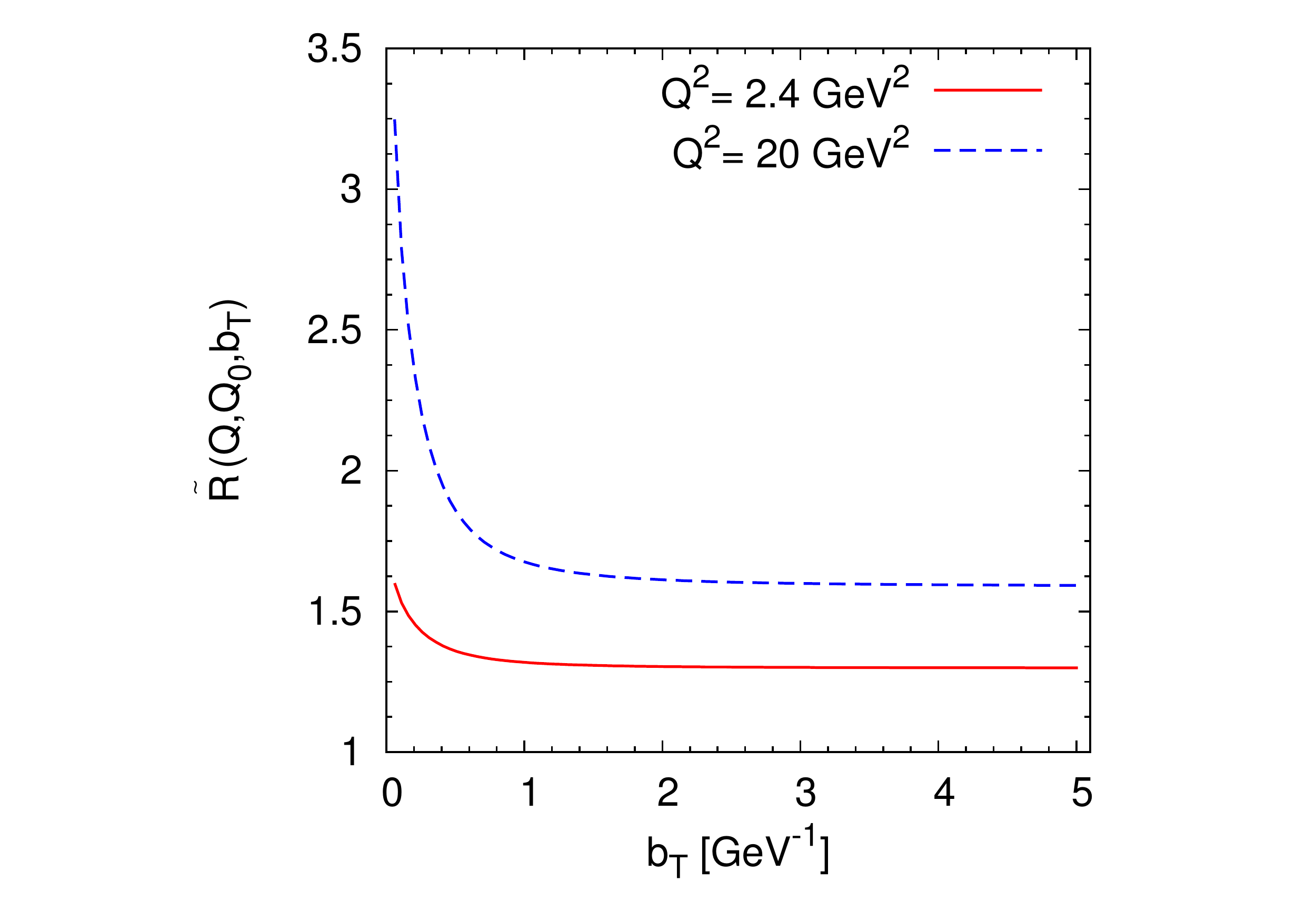}
\hspace*{0.5cm}
\includegraphics[width=0.3\textwidth,angle=-90]
{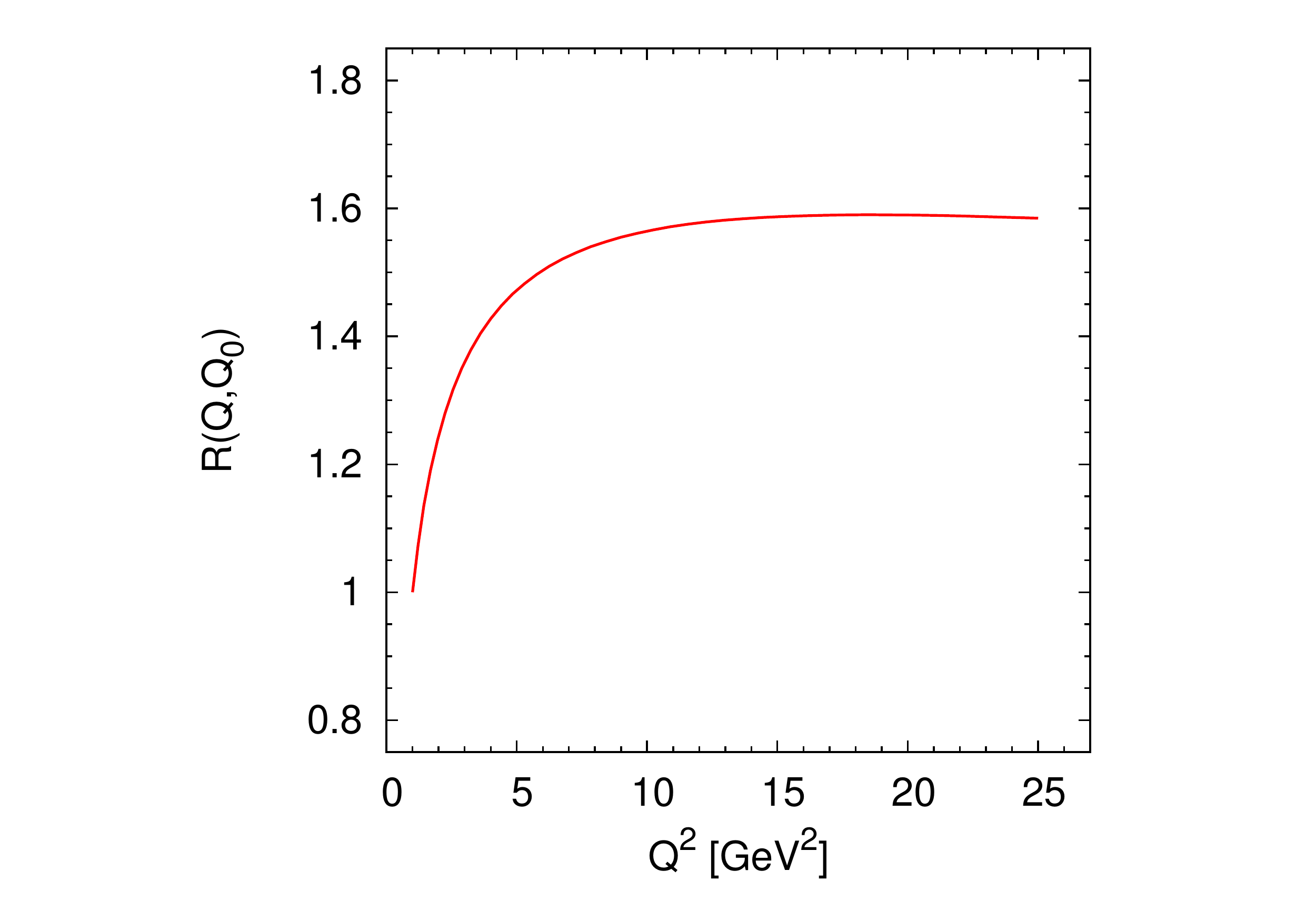}
\caption{\label{fig:R} In the left panel, the evolution factor $R(Q,Q_0,\bt)$ is plotted as a function of $\bt$ at two 
fixed values of $Q^2$. In the right panel we show $R(Q,Q_0)\equiv R(Q,Q_0,\bt \to \infty)$ as a function of $Q^2$. 
In both cases, $Q_0^2=1$ GeV$^2$.
}
\end{center}
\end{figure}

The TMD evolution in  Eqs.~(\ref{evF-f})-(\ref{evF-Sivers}) implies, 
apart from the explicit Gaussian dependence, a further non trivial 
dependence on the parton impact parameter $b_T$ through the evolution kernel 
$\widetilde R(Q,Q_0,\bt)$ and the upper integration limit $\mu_b$, 
Eq.~(\ref{mub}), which appears in Eq.~(\ref{RQQ0}); consequently, it needs 
to be evaluated numerically. However, the evolution equations can be solved 
analytically by making a simple approximation on this $b_T$ dependence. 
A close examination of Eq.~(\ref{mub}) shows that $\mu_b$ is a decreasing 
function of $b_T$ that very rapidly freezes to the constant value 
$C_1/b_{\rm max}=\mu_b(\bt \to \infty)$: more precisely, the approximation 
$\mu_b = const.$ holds for any $b_T \gsim 1$ GeV$^{-1}$.
As very small values of $\bt$ correspond to very large values of $\kt$, 
this approximation is safe in our framework, where the typical 
$\kt$ are less than $1$ GeV. Neglecting the $b_T$-dependence of $\mu_b$, 
the factor $\widetilde R(Q,Q_0,\bt)$ does not depend on $b_T$ anymore, see 
Eq.~(\ref{RQQ0}), and can even be integrated analytically by using an
explicit representation of $\alpha_s(Q)$. In the sequel we will refer to 
it as $R(Q,Q_0)$, with $R(Q,Q_0) \equiv \widetilde R(Q,Q_0,\bt\to \infty)$.
Fig.~\ref{fig:R} shows the evolution factor $\widetilde R(Q,Q_0,\bt)$ 
plotted as a function of $\bt$ at two fixed values of $Q^2$ (left panel), 
and $R(Q,Q_0)$ as a function of $Q^2$ (right panel). It is clear that 
$R(Q,Q_0)$ settles to a constant value for $b_T \gsim 1$ GeV$^{-1}$. 
In both cases, $Q_0^2=1$ GeV$^2$. 

Thus, in this approximation, the TMD evolution equation (\ref{Ftev}) only 
depends on $b_T$ through the non-perturbative function $g_K(b_T)$, which 
has been chosen to be a quadratic function of $b_T$, Eq.~(\ref{gk}), and 
through the $b_T$ dependence of the initial input function 
$\widetilde F(x, \bbt; Q_0)$ which has been chosen to be Gaussian. It results 
in a $b_T$-Gaussian form, with a width which depends logarithmically on 
$Q/Q_0$, for the TMD evolution equation. For the unpolarized TMD PDFs one has
\be
\widetilde f_{q/p}(x, \bbt; Q) =  f_{q/p}(x,Q_0) \; R(Q, Q_0)\;
\exp \left\{-\frac{b_T^2}{4} \left( \avkt  + 2 \, g_2 
\ln \frac{Q}{Q_0}\right) \right\} \>.
\label{evF-gauss}
\ee
%
%
%
Its Fourier-transform, Eqs.~(\ref{TMDunpf}), delivers a Gaussian distribution 
in the transverse momentum space as well:
\be
\widehat f_{q/p}(x,\kt;Q)=f_{q/p}(x,Q_0)\; R(Q,Q_0) \; 
\frac{ e^{-\kt ^2/w^2}}{\pi\,w^2} \>,\label{unp-gauss-evol} \quad\quad
\ee
where $f_{q/p}(x,Q_0)$ is the usual integrated PDF evaluated at the initial 
scale $Q_0$ and, most importantly, $w^2 \equiv w^2(Q,Q_0)$ is the ``evolving'' 
Gaussian width, defined as:
\be
w^2(Q,Q_0)=\avkt + 2\,g_2 \ln \frac{Q}{Q_0}\> \cdot \label{wf}
\ee 
It is worth noticing that the $Q^2$ evolution of the TMD PDFs is now 
determined by the overall factor $R(Q,Q_0)$ and, most crucially, by the 
$Q^2$ dependent Gaussian width $w(Q,Q_0)$. 

The TMD FFs evolve in a similar way, Eq.~(\ref{evF-D}), 
\be
\widetilde D_{h/q}(z, \bt; Q) = \frac{1}{z^2} D_{h/q}(z,Q_0) \; R(Q, Q_0)\;
\exp \left\{- \frac{b_T^2}{4 \, z^2} \left( \avpt \,  + 2 \, z^2 \, g_2 
\ln \frac{Q}{Q_0}\right) \right\} \>,
\label{evD-gauss}
\ee
leading to the TMD FF in momentum space,
\be
\widehat D_{h/q}(z,\pp;Q) =  D_{h/q}(z,Q_0)\; R(Q,Q_0) \; 
\frac{e^{-\pp^2/w^2_{\!F}}}{\pi w^2 _{\!F}} \>,
\label{D-gauss-evol} 
\ee
with an evolving and $z$-dependent Gaussian width $w_{\!F}\equiv 
w_{\!F}(Q,Q_0)$ given by
\be
w _{\!F}^2\equiv w _{\!F}^2(Q,Q_0)=\avp + 2 z^2 g_2 \ln \frac{Q}{Q_0}\>\cdot
\label{wd}
\ee  

For the Sivers distribution function, by Fourier-transforming 
Eq.~(\ref{evF-Sivers}) (with $\widetilde R \to R$) as prescribed by 
Eq.~(\ref{TMDsiv}), we obtain [see also Eqs.~(\ref{Siv1}), (\ref{Siv2}), 
(\ref{gamma}) and (\ref{fiTp})]:
\be
\Delta^N \widehat f_{q/p^\uparrow}(x,\kt;Q)=\frac{\kt}{M_1}\,\sqrt{2 e}\,
\frac{\avkt _S^2}{\avkt}\,\Delta^N f_{q/p^\uparrow}(x,Q_0)\,R(Q,Q_0)\, 
\frac{e^{-\kt^2/w_S^2}}{\pi w_S^4} \>,\label{siv-gauss-evol}
\ee
with
\be
w^2_S(Q,Q_0)=\avkt _S + 2 g_2 \ln \frac{Q}{Q_0}\> \cdot \label{ws}
\ee

It is interesting to notice that the evolution factor $R(Q,Q_0)$, 
controlling the TMD evolution according to Eqs.~(\ref{unp-gauss-evol}), 
(\ref{D-gauss-evol}) and (\ref{siv-gauss-evol}) is the same for all functions 
(TMD PDFs, TMD FFs and Sivers ) and is flavor independent: consequently it will appear, squared, in both numerator and denominator of the Sivers azimuthal asymmetry and, approximately, cancel out. Therefore, we can safely conclude 
that most of the TMD evolution of azimuthal asymmetries is controlled by the logarithmic $Q$ dependence of the $\kt$ Gaussian widths $w^2(Q,Q_0)$,
Eqs.~(\ref{wf}), (\ref{wd}) and (\ref{ws}). We will come back to this in 
Section \ref{newfit}.
%
\begin{figure}[t]
\begin{center}\hspace*{-1cm}
\includegraphics[width=0.3\textwidth,angle=-90]
{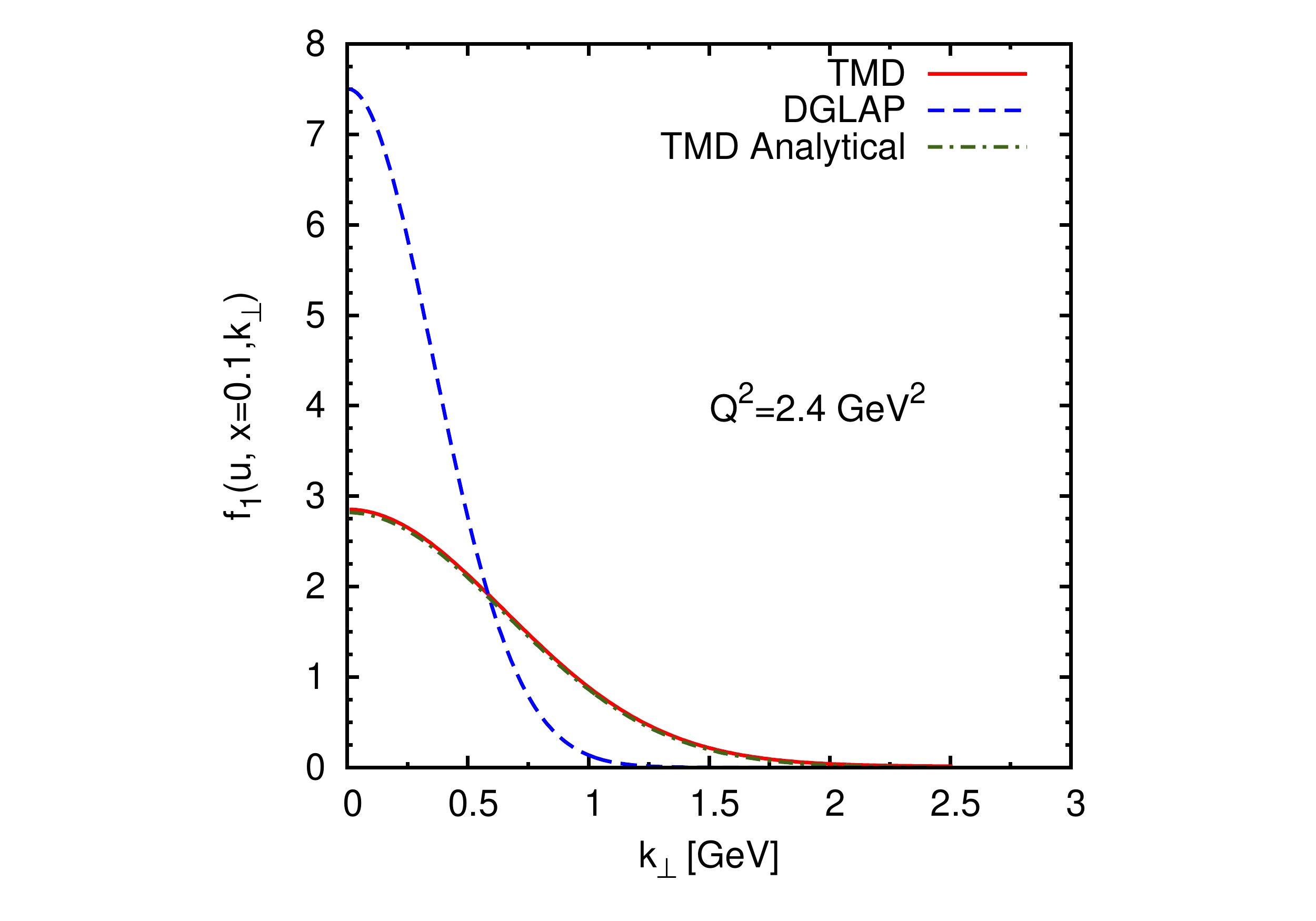}
\hspace*{-1cm}
\includegraphics[width=0.3\textwidth,angle=-90]
{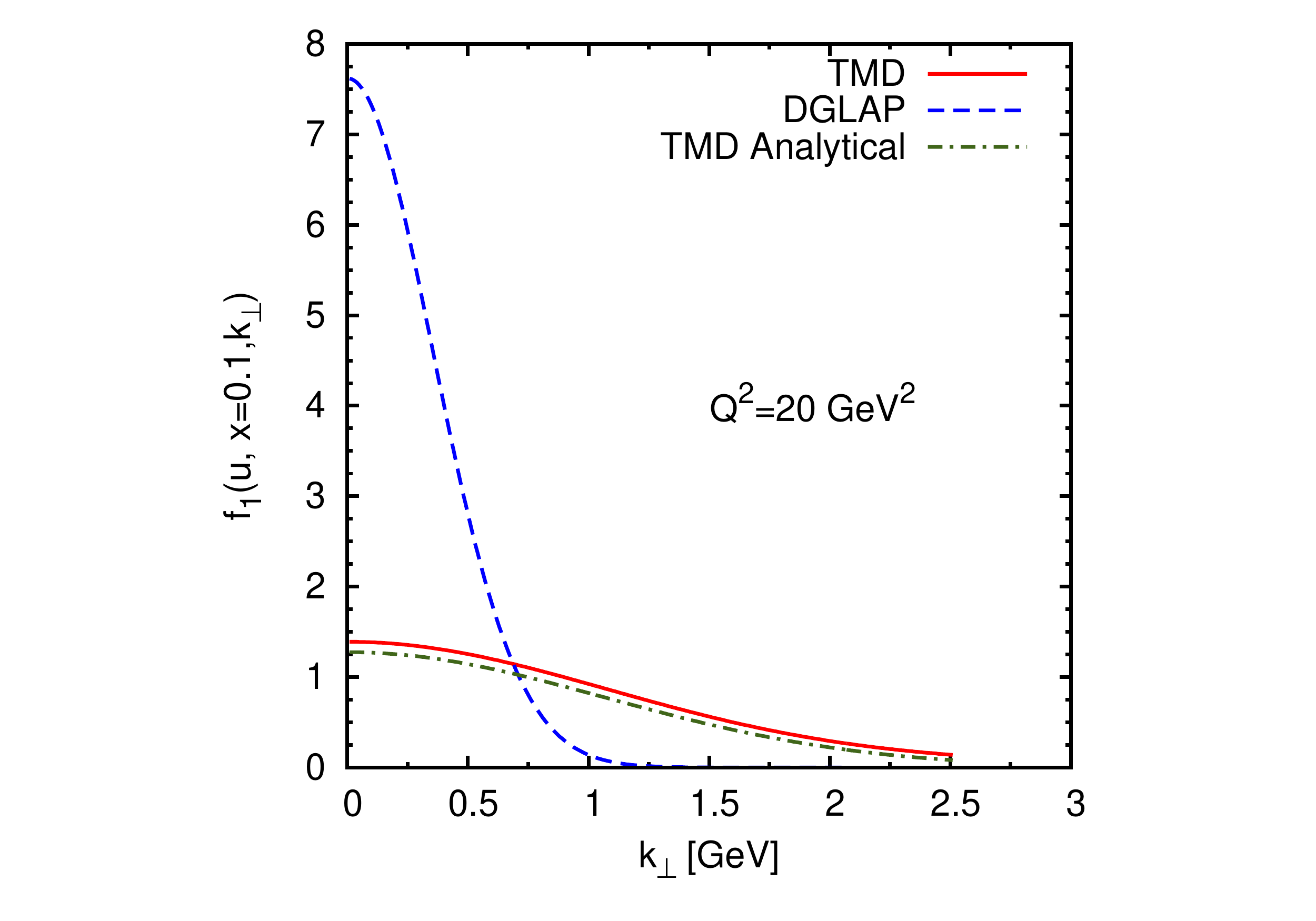}
\caption{\label{fig:unp-confronto} The left panel shows the unpolarized 
TMD PDF, $\widehat f_{u/p}$, evolved from the initial scale, $Q_0^2=1$ GeV$^2$, 
to $Q^2=2.4$ GeV$^2$, using TMD-evolution (red, solid line), DGLAP-evolution 
(blue, dashed line) and the analytical approximated TMD-evolution (green 
dot-dashed line). The right panel shows the same functions at the scale
$Q^2=20$ GeV$^2$. Notice that, while there is hardly any difference
between the DGLAP-evolved lines at $Q^2=2.4$ and $Q^2=20$ GeV$^2$, the TMD 
evolution induces a fast decrease in size of the TMD PDF functions at large 
$Q^2$ and a simultaneous widening of its Gaussian width. Here the 
analytical approximated evolution gives results in good agreement with 
the exact calculation even at large $Q^2$.  
}
\end{center}
\end{figure}
%
\begin{figure}[t]
\begin{center}\hspace*{-1cm}
\includegraphics[width=0.3\textwidth,angle=-90]
{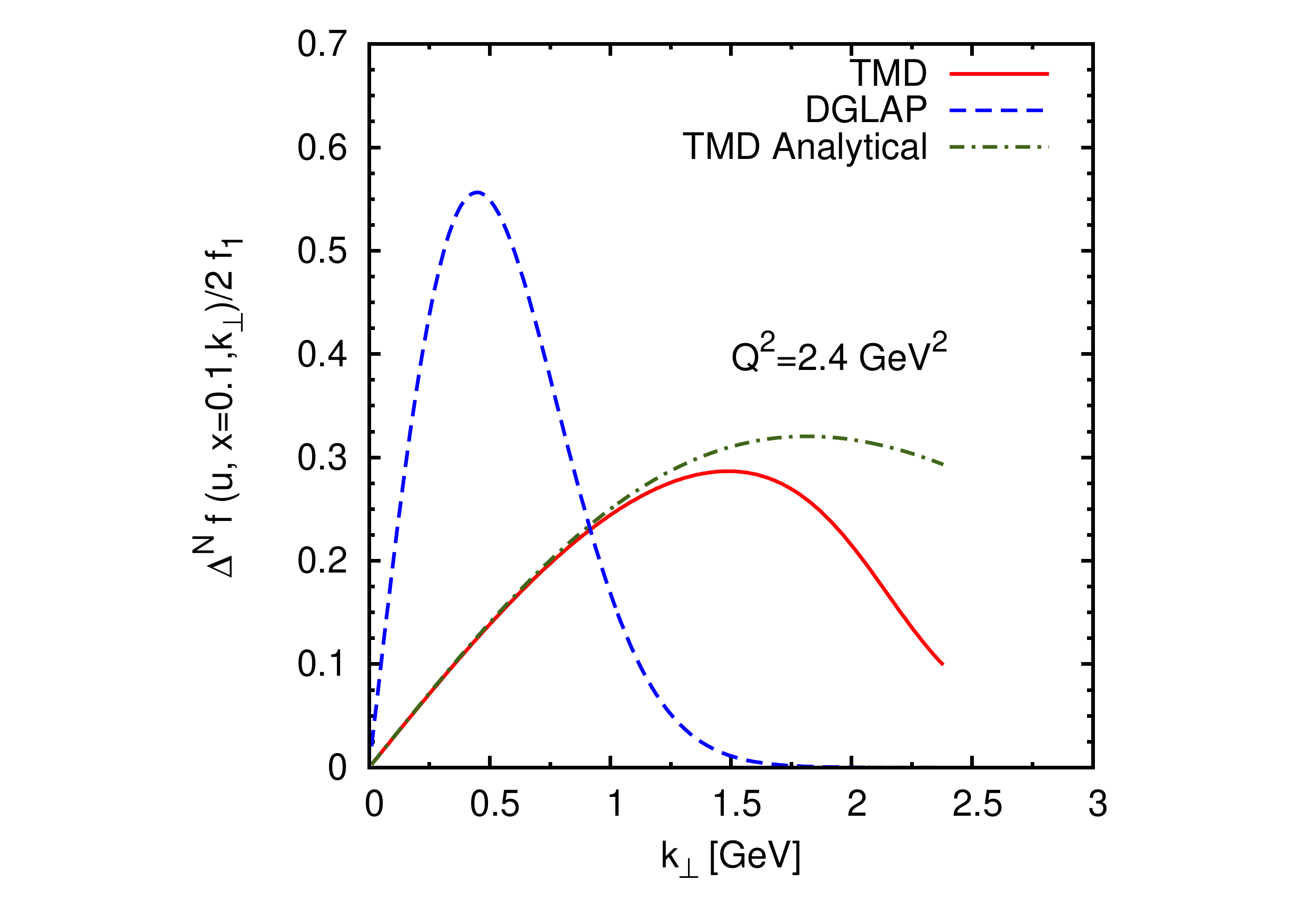}
\hspace*{-1cm}
\includegraphics[width=0.3\textwidth,angle=-90]
{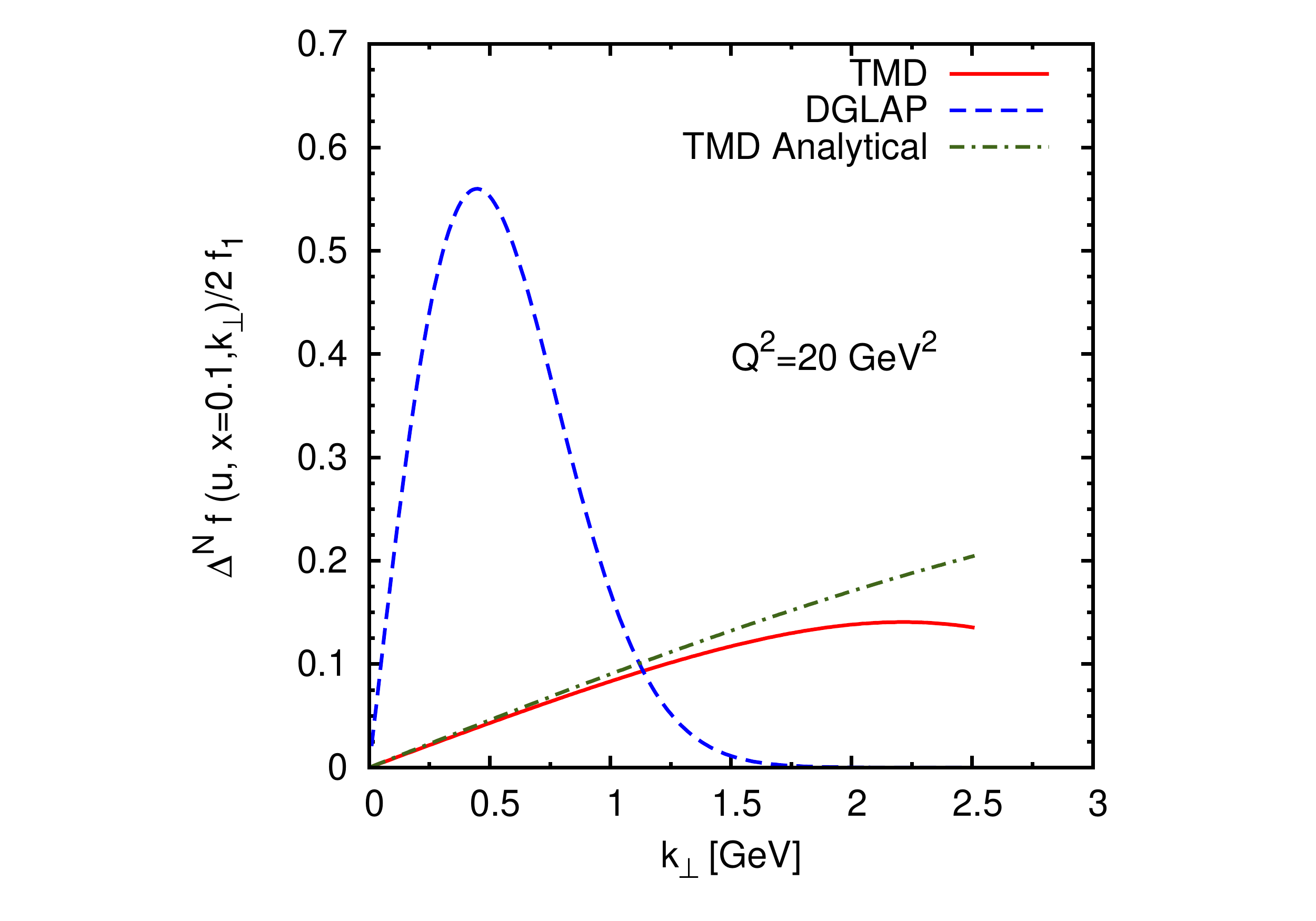}
\caption{\label{fig:siv-confronto} The left panel shows the ratio Sivers/PDF, 
$\Delta^N \widehat f_{u/p^{\uparrow}}/ 2 \widehat f_{u/p}$, evolved from the 
initial scale, $Q_0^2=1$ GeV$^2$, to $Q^2=2.4$ GeV$^2$, using TMD-evolution 
(red, solid line), DGLAP-evolution (blue, dashed line) and the analytical 
approximated TMD-evolution (green dot-dashed line). The right panel shows the 
same functions at the scale $Q^2=20$ GeV$^2$. Notice that, while there is 
almost no difference between the DGLAP-evolved lines at $Q^2=2.4$ and $Q^2=20$ 
GeV$^2$, the TMD evolution induces a fast decrease in size of the ratio 
Sivers/PDF functions with growing $Q^2$ and a simultaneous widening of
its Gaussian width. It is interesting to point out that the analytical 
approximation, for the Sivers function, visibly breaks down at large values 
of $Q^2$.
}
\end{center}
\end{figure}

To illustrate the features of this new TMD evolution, we compare it with the results obtained evolving only the collinear part, $f_{q/p}(x,Q)$, of the unpolarized TMD PDF according to the usual DGLAP equations and assuming the 
$\kt$ dependent term of this function to be unaffected by evolution. 
In the left panel of Fig.~\ref{fig:unp-confronto} we show the $\kt$ behavior 
of the unpolarized TMD PDF $\widehat f_{u/p}(x,\kt,Q^2)$, at the fixed value 
$x=0.1$, evaluated at the scale $Q^2=2.4$ GeV$^2$ (the average $Q^2$ value for 
the HERMES experiment). In the right panel we show the same function at a higher 
scale, $Q^2=20$ GeV$^2$ (which is the highest bin average $Q^2$ detected in 
the COMPASS experiment). In both cases the chosen initial scale is $Q_0^2=1$ 
GeV$^2$. The red, solid line corresponds to the $\kt$ distribution 
of the TMD PDF found by using the TMD-evolution of 
Eq.~(\ref{evF-f}) while the blue, dashed line represents the result obtained 
by using DGLAP evolution equations. At the initial scale, $Q_0^2=1$ GeV$^2$, 
solid and dashed curves coincide, by definition. However, while the DGLAP 
evolution is so slow that  there is hardly any difference between the 
DGLAP-evolved lines at $Q^2=2.4$ GeV$^2$ and $Q^2=20$ GeV$^2$, the TMD 
evolution induces a fast decrease of the maximum values of the TMD PDF 
function with growing $Q^2$, and a simultaneous broadening of its Gaussian 
width, as observed in Refs.~\cite{Aybat:2011zv} and~\cite{Aybat:2011ge}. 
It is interesting to notice that the approximated evolution of 
Eq.~(\ref{unp-gauss-evol}), corresponding to the green, dot-dashed line works really well, even for large $Q^2$ values.  

A similar study is performed in Fig.~\ref{fig:siv-confronto} for the Sivers 
function. Here, by DGLAP evolution we mean that the Sivers function evolves 
like an unpolarized collinear PDF, only through the factor $f_{q/p}(x,Q)$ 
contained in its parameterization, Eq.~(\ref{DeltaN}). The parameters used 
for the plots are those given in Table~\ref{tab:fitpar_sivers-tmd}, although 
any set of realistic parameters would lead to the same conclusions. The left 
panel shows the ratio between the Sivers function and the TMD PDF, $\Delta^N 
\widehat f_{u/p^{\uparrow}}(x,\kt;Q)/ (2 \widehat f_{u/p}(x,\kt;Q))$, 
evaluated at the scale $Q^2=2.4$ GeV$^2$. Again, the red, solid line is 
obtained using the TMD-evolution of Eqs.~(\ref{evF-f}) and (\ref{evF-Sivers}), while the blue, dashed line is given by the DGLAP-evolution. The green 
dot-dashed line represents the results obtained using the approximated 
analytical TMD-evolution of Eqs.~(\ref{unp-gauss-evol}) and 
(\ref{siv-gauss-evol}). The right panel shows the same functions at the 
scale $Q^2=20$ GeV$^2$. Similarly to the case of TMD PDFs, while there is no 
difference between the DGLAP-evolved lines at $Q^2=2.4$ and $Q^2=20$ GeV$^2$, 
the TMD evolution induces a fast decrease in the size of the TMD Sivers 
functions with growing $Q^2$ and a simultaneous widening of its Gaussian 
width. It is interesting to point out that the analytical TMD approximation, 
for the Sivers function visibly breaks down for large values of $\kt$.

\section{\label{newfit} SIDIS data and TMD vs. non-TMD evolution} 

Having established the phenomenological formalism necessary to implement the 
TMD evolution, as given in Refs.~\cite{Collins:2011book, Aybat:2011zv,
Aybat:2011ge}, we apply it to the Sivers function. This TMD distribution, 
$\Delta^N \widehat f_ {q/\pup}(x,k_\perp,Q) = (-2\kt/M_p)\widehat f_{1T}^\perp$, 
can be extracted from HERMES and COMPASS $\ell \, p \to h \, X$ SIDIS data on 
the azimuthal moment $A^{\sin(\phi_h-\phi_S)}_{UT}$, defined as
\be
A^{\sin (\phi_h-\phi_S)}_{UT} = \label{def-siv-asym}
2 \, \frac{\int d\phi_S \, d\phi_h \,
[d\sigma^\uparrow - d\sigma^\downarrow] \, \sin(\phi_h-\phi_S)}
{\int d\phi_S \, d\phi_h \,
[d\sigma^\uparrow + d\sigma^\downarrow]}\, \cdot
\ee
This transverse single spin asymmetry (SSA) embeds the azimuthal modulation triggered by the correlation between the nucleon spin and the quark intrinsic transverse momentum. The ``weighting'' factor $\sin(\phi_h -\phi_S)$ in 
Eq.~(\ref{def-siv-asym}) is appropriately chosen to single out, among the 
various azimuthal dependent terms appearing in 
$[d\sigma^\uparrow - d\sigma^\downarrow]$, only the contribution of the Sivers
mechanism~\cite{Anselmino:2011ch,Bacchetta:2006tn}. 
By properly taking into account all intrinsic motions this
transverse single spin asymmetry can be written as \cite{Anselmino:2005ea}
\be
A^{\sin (\phi_h-\phi_S)}_{UT} = \label{hermesut}
\frac{\displaystyle  \sum_q \int
{d\phi_S \, d\phi_h \, d^2 \bfk _\perp}\;
\Delta^N \widehat f_{q/\pup} (x,\kt,Q) \; \sin (\varphi -\phi_S) \;
\frac{d \hat\sigma ^{\ell q\to \ell q}}{dQ^2} 
\; \widehat D_q^h(z,p_\perp,Q) \sin (\phi_h -\phi_S) }
{\displaystyle \sum_q \int {d\phi_S \,d\phi_h \, d^2 \bfk _\perp}\;
\widehat f_{q/p}(x,k _\perp,Q) \; \frac{d \hat\sigma ^{\ell q\to \ell q}}
{dQ^2} \; \widehat D_q^h(z,p_\perp,Q) } \> \cdot
\ee
With respect to the leptonic plane, $\phi_S$ and $\phi_h$ are the azimuthal 
angles identifying the transverse directions of the proton spin $\bfS$ and 
of the outgoing hadron $h$ respectively, while $\varphi$ defines the 
direction of the incoming (and outgoing) quark transverse momentum,
$\bfk_\perp$ = $\kt(\cos\varphi, \sin\varphi,0)$;
${d \hat\sigma ^{\ell q\to \ell q}}/{dQ^2}$ is the unpolarized
cross section for the elementary scattering  $\ell q\to \ell q$.
 
The aim of our paper is to analyze the available polarized SIDIS data from 
the HERMES and COMPASS collaborations in order to understand whether or not 
they show signs of the TMD evolution proposed in Ref.~\cite{Aybat:2011ge} 
and described in Section~\ref{Intro-sub}. Our general strategy is that of 
adopting the TMD evolution in the extraction of the Sivers functions, with 
the same parameterization and input functions as in 
Refs.~\cite{Anselmino:2008sga,Anselmino:2011gs}, and see if that 
can improve the quality of the fits. In doing so we will make use of the 
HERMES re-analysis of SIDIS experimental data on Sivers asymmetries for pion
and kaon production and the newest SIDIS COMPASS data off a proton 
target, which cover a wider range of $Q^2$ values, thus giving a better 
opportunity to check the TMD evolution.

In particular we perform three different data fits:
\begin{itemize} 
\item a fit (TMD-fit) in which we adopt the TMD evolution equation 
discussed in the Section~\ref{Intro-sub} and \ref{Param}, 
Eqs.~(\ref{evF-f})-(\ref{evF-Sivers}) and (\ref{TMDunpf})-(\ref{TMDsiv});
\item  a second fit (TMD-analytical-fit) in which we apply the same TMD evolution, 
but using the analytical approximation discussed in Section~\ref{gaussian}, Eqs.~(\ref{unp-gauss-evol}), (\ref{D-gauss-evol}) and (\ref{siv-gauss-evol});
\item a fit (DGLAP-fit) in which we follow our previous work, as done so far 
in Ref.~\cite{Anselmino:2008sga,Anselmino:2011gs}, using the DGLAP evolution 
equation only in the collinear part of the TMDs. 
\end{itemize}
As a result of the fit we will have explicit expressions of all the Sivers 
functions and their parameters. However, the goal of the paper is not that 
of obtaining a new extraction of the Sivers distributions, although in the 
sequel we will show, for comment and illustration purposes, the Sivers 
functions for $u$ and $d$ valence quarks, with the relative parameters. 
The procedure followed here aims at testing the effect of the TMD evolution, 
as compared with the simple DGLAP evolution so far adopted, in fitting the 
TMD SIDIS data. If it turns out, as it will, that this improves the quality 
of the fit, then a new extraction of the Sivers distributions, entirely 
guided by the TMD evolution, will be necessary. That will require a different 
approach from the very beginning, with different input functions and 
parameterizations. 

Here, we parameterize the Sivers function at the initial scale $Q_0=1$ GeV,
as in Ref.~\cite{Anselmino:2008sga,Anselmino:2011gs}, in the following form:
\be
\Delta^N \widehat f_ {q/\pup}(x,\kt,Q_0) = \Delta^N \! f_ {q/\pup}(x,Q_0)\, h(\kt) 
= 2 \, {\cal N}_q(x) \, h(\kt) \,
\widehat f_ {q/p} (x,\kt,Q_0)\; , \label{sivfac}
\ee
with
\bea
&&{\cal N}_q(x) =  N_q \, x^{\alpha_q}(1-x)^{\beta_q} \,
\frac{(\alpha_q+\beta_q)^{(\alpha_q+\beta_q)}}
{\alpha_q^{\alpha_q} \beta_q^{\beta_q}}\; ,
\label{siversx} \\
&&h(\kt) = \sqrt{2e}\,\frac{k_\perp}{M_{1}}\,e^{-{k_\perp^2}/{M_{1}^2}}\; ,
\label{siverskt}
\eea
where $\widehat f_ {q/p} (x,\kt,Q_0)$ is defined in Eq.~\ref{partonf} and
$N_q$, $\alpha_q$, $\beta_q$ and $M_1$ (GeV) are (scale independent) free parameters to be determined by fitting the experimental data. Since 
$h(\kt) \le 1$ for any $\kt$ and $|{\cal N}_q(x)| \le 1$ for any $x$ 
(notice that we allow the constant parameter $N_q$ to vary only inside the 
range $[-1,1]$), the positivity bound for the Sivers function,
\be
\frac{|\Delta^N\widehat f_ {q/\pup}(x,\kt)|}
{2 \widehat f_ {q/p} (x,\kt)}\le 1\>,
\label{pos}
\ee
is automatically fulfilled. Similarly to PDFs, the FFs at the initial scale 
are parameterized with a Gaussian shape, Eq.~(\ref{partond}).

As in Refs.~\cite{Anselmino:2005nn} and~\cite{Anselmino:2008sga}, the average values of $\kt$ and $\pp$ are fixed as
\be
\langle\kt^2\rangle   = 0.25  \;{\rm GeV}^2 \quad\quad\quad
\langle p_\perp^2\rangle  = 0.20 \;{\rm GeV}^2 \>.
\label{ktpar}
\ee

We take the unpolarized distributions $f_{q/p}(x,Q^2_0)$ from 
Ref.~\cite{Gluck:1998xa} and the unpolarized fragmentation functions 
$D_{h/q}(z,Q^2_0)$ from Ref.~\cite{deFlorian:2007aj}, with $Q^2_0=1.0$ GeV.
As in Ref.~\cite{Anselmino:2008sga}, we adopt 11 free parameters:
\bea
&& N_{u_v} \quad\quad\quad  N_{d_v} \quad\quad\quad  N_s
\nonumber \\
&& N_{\bar u} \quad\quad\quad\;\,  N_{\bar d} \quad\quad\quad\;  N_{\bar s}
\nonumber \\
&& \alpha_{u_v} \quad\quad\quad \,\alpha_{d_v} \quad\quad\quad \,\alpha_{sea} \label{par_broken}\\
&& \beta   \quad\quad\quad\,\,\,\;\; M_1\;({\rm GeV}) \>, \nonumber
\eea
where the subscript $v$ denotes valence contributions. In this choice we differ from Ref.~\cite{Anselmino:2008sga}, where valence and sea contributions were not separated.

We perform best fits of $11$ experimental data sets: HERMES~\cite{:2009ti} 
data for SIDIS production of pions ($\pi^{+}$, $\pi^{-}$, $\pi^{0}$) and kaons 
($K^{+}$ and $K^{-}$), COMPASS data for SIDIS pion ($\pi^{+}$, $\pi^{-}$) and 
kaon ($K^{+}$ and $K^{-}$) production from a $LiD$ (deuteron) 
target~\cite{:2008dn}, and the preliminary COMPASS data for charged hadron 
production from an $NH_3$ (proton) target~\cite{Bradamante:2011xu}. The results 
of these 3 fits are presented in Table~\ref{tab:chi-sq} in terms of their 
$\chi^2$s. 

\begin{table}[t]
\caption{$\chi^2$ contributions corresponding to the TMD-fit,
the TMD-analytical-fit 
and the DGLAP-fit, for each experimental data set of HERMES and 
COMPASS experiments.\label{tab:chi-sq}}
\vspace*{6pt}
\begin{ruledtabular}
\begin{tabular}{cccccc}
\noalign{\vspace{3pt}}
&  &   & TMD Evolution (exact)   & TMD Evolution (analytical) & DGLAP Evolution\\
\noalign{\vspace{3pt}}
\cline{4-6}
\noalign{\vspace{3pt}}
&  &   & $\chi^2_{tot}\;=255.8$   & $\chi^2_{tot}\;=275.7$    &
$\chi^2_{tot}\;=315.6$\\
\noalign{\vspace{1pt}}
&  &   & $\chi^2_{d.o.f}=\;1.02$ & $\chi^2_{d.o.f}=\;1.10$  & $\chi^2_{d.o.f}=\;1.26$\\
\noalign{\vspace{5pt}}
\noalign{\vspace{3pt}}
 Experiment & Hadron  & N. points & & &\\
\noalign{\vspace{3pt}}
\cline{1-6}
\noalign{\vspace{3pt}}
                  &            & 7 & $\chi^2_x\;=10.7$      &  $\chi^2_x\;=12.9$    & $\chi^2_x\;=27.5$   \\
                  &  $\pi^+$   & 7 & $\chi^2_z\;=\;4.3$     &  $\chi^2_z\;=\;4.3$   & $\chi^2_z\;=8.6$     \\
                  &            & 7 & $\chi^2_{P_T}\!=9.1$   &  $\chi^2_{P_T}\!=10.5$& $\chi^2_{P_T}\!=22.5$ \\
\noalign{\vspace{3pt} }
\cline{2-6}
\noalign{\vspace{3pt}}
                  &            & 7 & $\chi^2_x=17.0$      &  $\chi^2_x=16.5$     &$\chi^2_x=14.8$   \\
                  &  $\pi^-$   & 7 & $\chi^2_z=\;2.4$     &  $\chi^2_z=\;2.4$    &$\chi^2_z=\;3.3$   \\
                  &            & 7 & $\chi^2_{P_T}\!=6.4$   &  $\chi^2_{P_T}\!=6.3$ & $\chi^2_{P_T}\!=6.2$  \\
\noalign{\vspace{3pt}}
\cline{2-6}
\noalign{\vspace{3pt}}
                  &            & 7 & $\chi^2_x=\;5.9$     &  $\chi^2_x=\;5.8$    &  $\chi^2_x=\;5.6$ \\
  HERMES          &  $\pi^0$   & 7 & $\chi^2_z=\;8.0$     &  $\chi^2_z=\;8.1$    &  $\chi^2_z=\;6.9$  \\
                  &            & 7 & $\chi^2_{P_T}\!=6.8$   &  $\chi^2_{P_T}\!=7.0$ & $\chi^2_{P_T}\!=6.6$   \\
\noalign{\vspace{3pt}}
\cline{2-6}
\noalign{\vspace{3pt}}
                  &            & 7 & $\chi^2_x=\;4.7$     &  $\chi^2_x=\;4.8$    & $\chi^2_x=\;4.4$   \\
                  &  $K^+$     & 7 & $\chi^2_z=\;9.3$     &  $\chi^2_z=\;9.8$    & $\chi^2_z=\;4.3$    \\
                  &            & 7 & $\chi^2_{P_T}\!=4.6$ &  $\chi^2_{P_T}\!=5.3$  & $\chi^2_{P_T}\!=2.8$ \\
\noalign{\vspace{3pt}}
\cline{2-6}
\noalign{\vspace{3pt}}
                  &            & 7 & $\chi^2_x=\;2.4$     &  $\chi^2_x=\;2.4$    & $\chi^2_x=\;2.9$   \\
                  &  $K^-$     & 7 & $\chi^2_z=\;7.2$     &  $\chi^2_z=\;7.0$    & $\chi^2_z=\;5.5$    \\
                  &            & 7 & $\chi^2_{P_T}\!=3.4$ &  $\chi^2_{P_T}\!=3.3$  & $\chi^2_{P_T}\!=3.7$ \\
\noalign{\vspace{3pt}}
\cline{1-6}
\noalign{\vspace{3pt}}
                  &            & 9 & $\chi^2_x=\;6.7$     &  $\chi^2_x=11.2$     & $\chi^2_x=29.2$   \\
                  &  $h^+$     & 8 & $\chi^2_z=17.8$      &  $\chi^2_z=18.5$     & $\chi^2_z=16.6$   \\
COMPASS-p         &            & 9 & $\chi^2_{P_T}\!=12.4$ &  $\chi^2_{P_T}\!=24.2$&  $\chi^2_{P_T}\!=11.8$   \\
\noalign{\vspace{3pt}}
\cline{2-6}
\noalign{\vspace{3pt}}
                  &            & 9 & $\chi^2_x=\;7.6$     &  $\chi^2_x=\;7.7$    & $\chi^2_x=11.9$   \\
                  &  $h^-$     & 8 & $\chi^2_z=\;9.7$     &  $\chi^2_z=\;9.6$    & $\chi^2_z=14.1$    \\
                  &            & 9 & $\chi^2_{P_T}\!=8.1$   &  $\chi^2_{P_T}\!=8.1$ &  $\chi^2_{P_T}\!=9.9$  \\
\noalign{\vspace{3pt}}
\cline{1-6}
\noalign{\vspace{3pt}}
                  &            & 9 & $\chi^2_x=\;7.3$     &  $\chi^2_x=\;7.1$    & $\chi^2_x=\;5.3$  \\
                  &  $\pi^+$   & 8 & $\chi^2_z=\;5.4$     &  $\chi^2_z=\;5.3$    & $\chi^2_z=\;7.9$   \\
                  &            & 9 & $\chi^2_{P_T}\!=5.4$   &  $\chi^2_{P_T}\!=5.2$ &  $\chi^2_{P_T}\!=5.5$  \\
\noalign{\vspace{3pt}}
\cline{2-6}  
\noalign{\vspace{3pt}}
                  &            & 9 & $\chi^2_x=\;4.4$     &  $\chi^2_x=\;4.4$      & $\chi^2_x=\;5.0$  \\
                  &  $\pi^-$   & 8 & $\chi^2_z=10.9$      &  $\chi^2_z=10.7$       & $\chi^2_z= 13.9$   \\
COMPASS-d         &            & 9 & $\chi^2_{P_T}\!=4.5$ &  $\chi^2_{P_T}\!=4.8$  & $\chi^2_{P_T}\!=4.4$  \\
\noalign{\vspace{3pt}}
\cline{2-6}
\noalign{\vspace{3pt}}
                  &            & 9 & $\chi^2_x=\;6.5$      &  $\chi^2_x=\;6.5$      &  $\chi^2_x=\;5.8$   \\
                  &  $K^+$     & 8 & $\chi^2_z=\;7.7$      &  $\chi^2_z=\;7.7$      &  $\chi^2_z=\;7.2$    \\
                  &            & 9 & $\chi^2_{P_T}\!=4.8$  &  $\chi^2_{P_T}\!=4.9$  &  $\chi^2_{P_T}\!=4.7$   \\
\noalign{\vspace{3pt}}
\cline{2-6}
\noalign{\vspace{3pt}}
                  &            & 9 & $\chi^2_x=12.1$       &  $\chi^2_x=12.4$       &  $\chi^2_x=13.1$    \\
                  &  $K^-$     & 8 & $\chi^2_z=\;8.9$      &  $\chi^2_z=\;9.0$      &  $\chi^2_z=9.4$     \\
                  &            & 9 & $\chi^2_{P_T}\!=13.5$ &  $\chi^2_{P_T}\!=12.0$ &  $\chi^2_{P_T}\!=14.4$  \\
\noalign{\vspace{3pt}}
\end{tabular}
\end{ruledtabular}
\end{table}

As it is clear from the first line of Table~I, the best total $\chi^2_{tot}$, 
which amounts to $256$, is obtained by using the TMD evolution, followed 
by a slightly higher $\chi^2_{tot}$ of the analytical approximation, and 
a definitely larger $\chi^2_{tot} \simeq 316$ corresponding to the DGLAP fit.
To examine the origin of this difference between TMD and DGLAP 
evolution, we show the individual contributions to $\chi^2_{tot}$ of 
each experiment (HERMES, COMPASS on $NH_3$ and on $LiD$ targets), for all 
types of detected hadrons and for all variables observed ($x$, $z$ and $P_T$).
A global look at the numbers reported in Table~I shows that the difference 
of about 60 $\chi^2$-points between the TMD and the DGLAP fits is not equally distributed among all $\chi^2$s per data point; rather, it is heavily 
concentrated in three particular cases, namely in the asymmetry for 
$\pi^+$ production at HERMES and for $h^+$ and $h^-$ production at COMPASS 
off a proton target, especially when this asymmetry is observed as a function 
of the $x$-variable.

It is important to stress that, as $x$ is directly proportional to $Q^2$ 
through the kinematical relation $Q^2=x\,y\,s$, the $x$ behavior of the 
asymmetries is intimately connected to their $Q^2$ evolution. While the HERMES experimental bins cover a very modest range of $Q^2$ values, from $1.3$ GeV$^2$ 
to $6.2$ GeV$^2$, COMPASS data raise to a maximum $Q^2$ of $20.5$ GeV$^2$, 
enabling to test more severely the TMD $Q^2$ evolution in SIDIS.   

\begin{figure}[t]
\begin{center}\hspace*{-1cm}
\includegraphics[width=0.38\textwidth,angle=-90]
{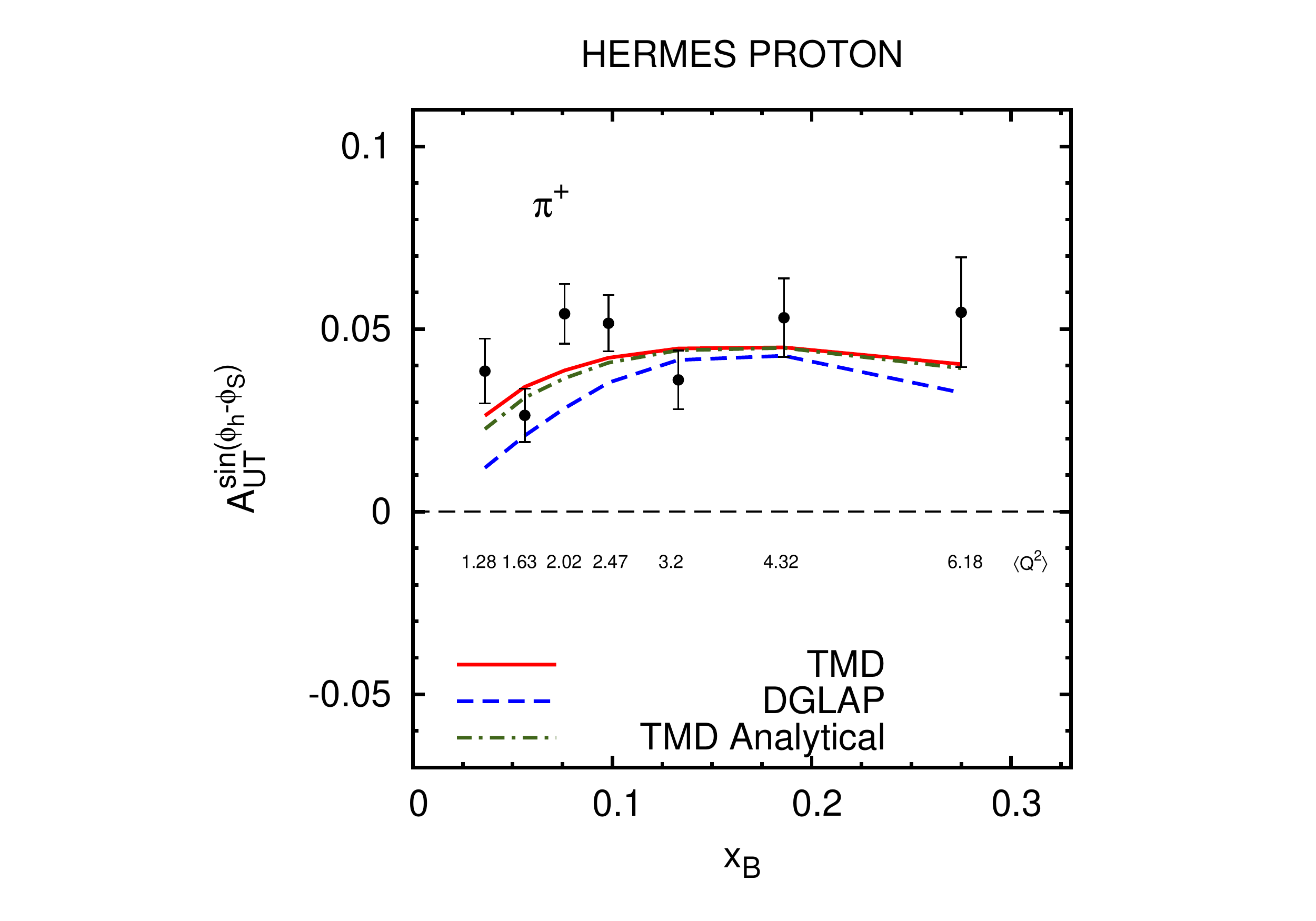}
\hspace*{-2cm}
\includegraphics[width=0.4\textwidth,angle=-90]
{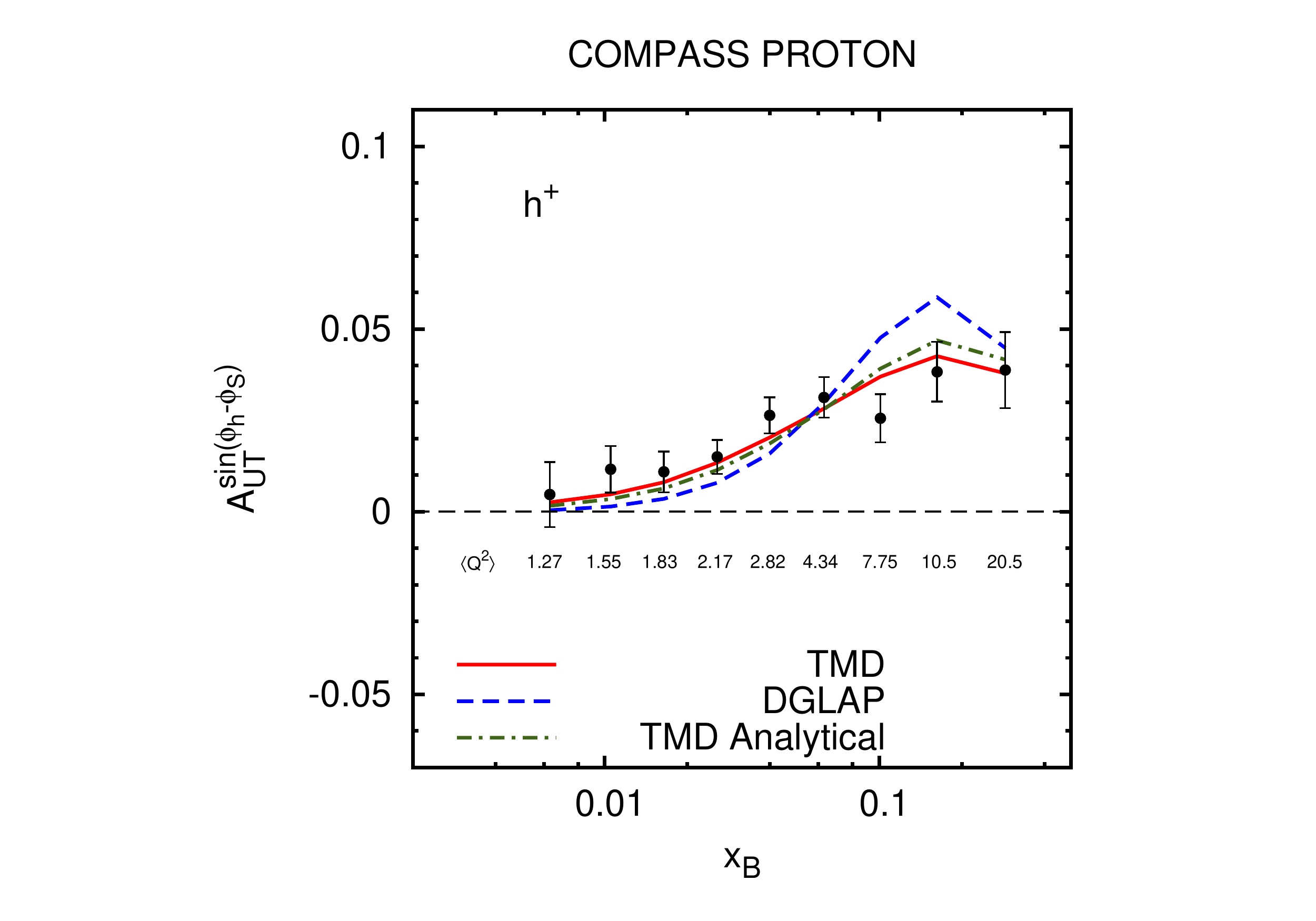}
\caption{\label{fig:hermes-compass-confronto}
The results obtained from our fit of the SIDIS $A_{UT}^{\sin{(\phi_h-\phi_S)}}$ Sivers asymmetries applying TMD evolution (red, solid lines) are compared with 
the analogous results found by using DGLAP evolution equations (blue, dashed lines). The green, dash-dotted lines correspond to the results obtained by 
using the approximated analytical TMD evolution (see text for further details). 
The experimental data are from HERMES~\cite{:2009ti} (left panel) and 
COMPASS~\cite{Bradamante:2011xu} (right panel) Collaborations. 
}
\end{center}
\end{figure}

These aspects are illustrated in Fig.~\ref{fig:hermes-compass-confronto}, 
where the SIDIS Sivers asymmetries  $A^{\sin(\phi_h-\phi_S)}_{UT}$ obtained 
in the three fits are shown in the same plot. It is evident that the DGLAP 
evolution seems to be unable to describe the correct $x$ trend, {\it i.e.} 
the right $Q^2$ behavior, while the TMD evolution (red solid line) follows 
much better the large $Q^2$ data points, corresponding to the last $x$-bins 
measured by COMPASS. The approximate analytical TMD evolution (green 
dash-dotted line) works very well for low to moderate values of $Q^2$ 
while it starts to deviate from the exact behavior at large $Q^2$ values. 
 
%
%
\begin{figure}[t]
\begin{center}
\includegraphics[width=0.43\textwidth, angle=-90]
{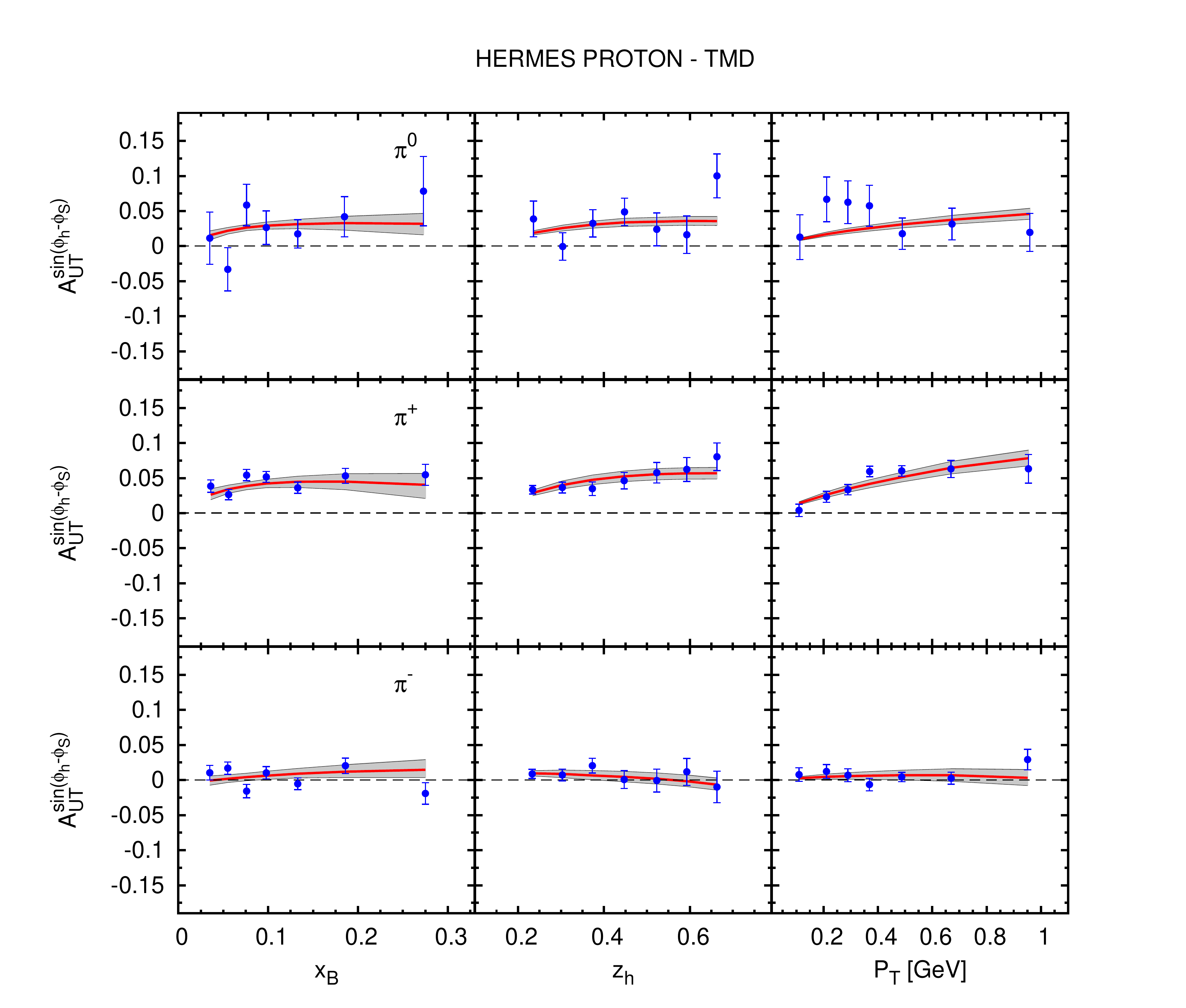}\hspace*{-.4cm}
\includegraphics[width=0.43\textwidth, angle=-90]
{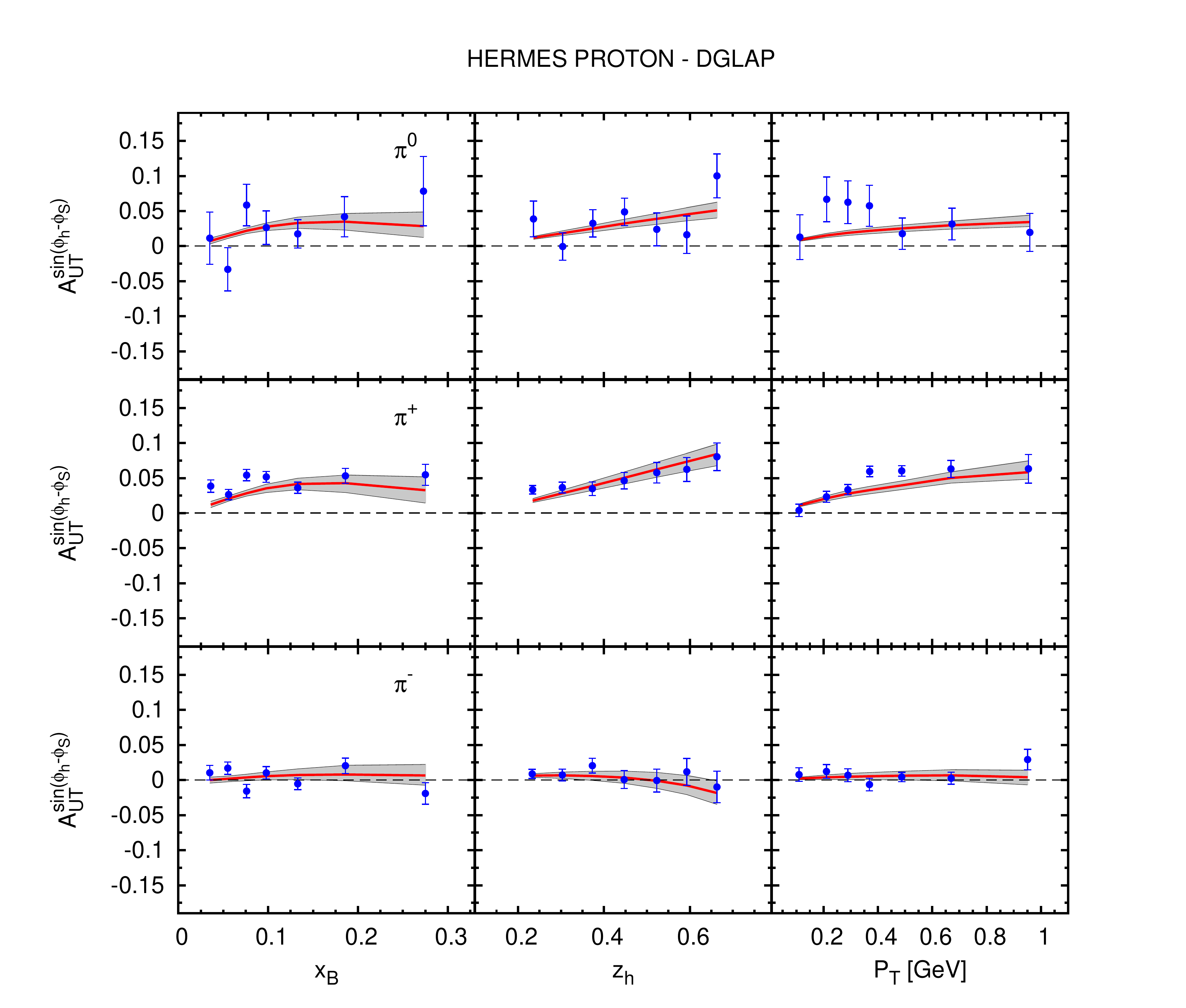}
\caption{\label{fig:hermes-pi}
The results obtained from the TMD-evolution fit (left panel) and from the  
DGLAP-evolution fit (right panel) of the SIDIS $A_{UT}^{\sin{(\phi_h-\phi_S)}}$ 
Sivers asymmetries (red, solid lines) are compared with the HERMES experimental 
data~\cite{:2009ti} for charged and neutral pion production. 
The shaded area corresponds to the statistical uncertainty of the parameters, 
see Appendix~A of Ref.~\cite{Anselmino:2008sga} for further details.
}
\end{center}
\end{figure}
%
\begin{figure}[t]
\begin{center}
\includegraphics[width=0.3\textwidth, angle=-90]
{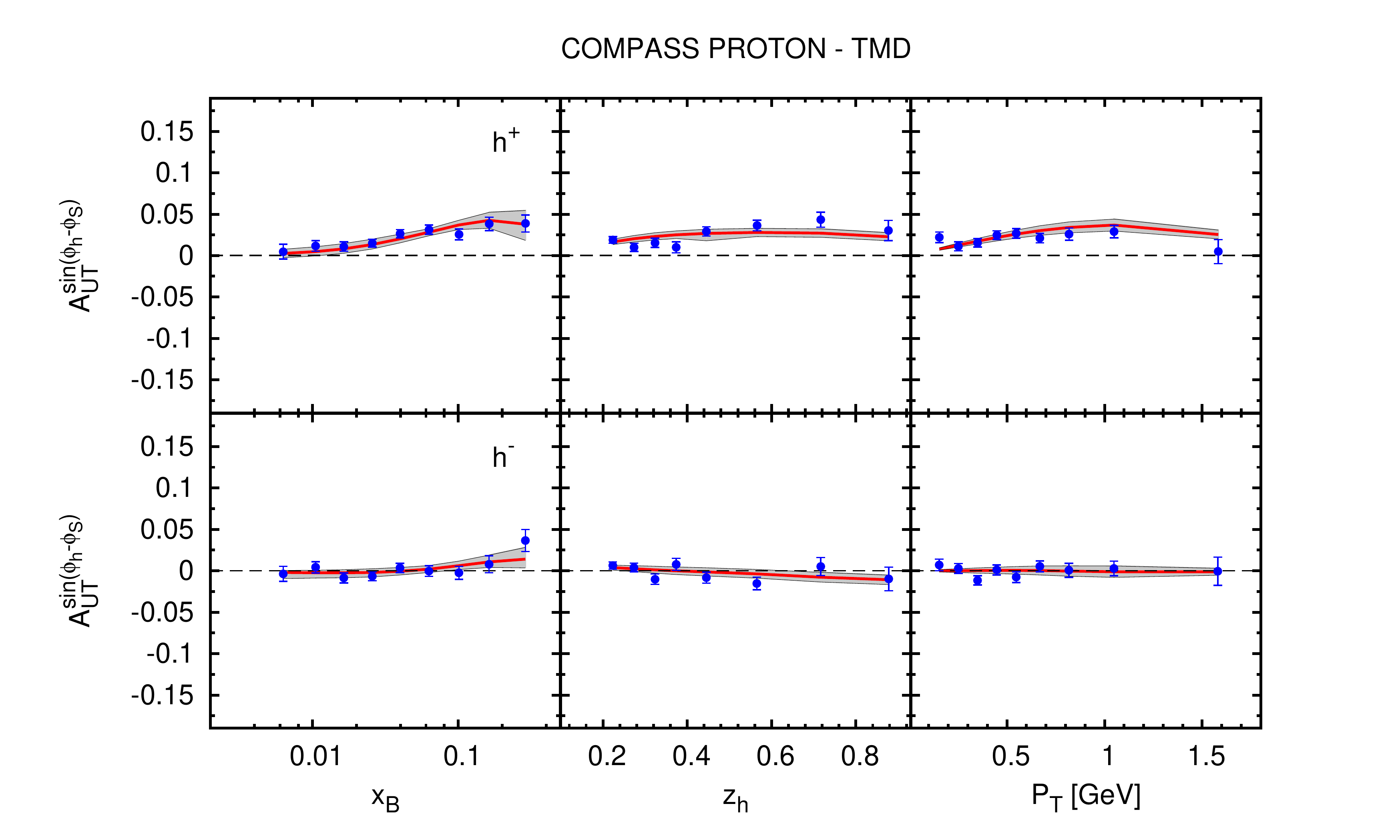}\hspace*{-.2cm}
\includegraphics[width=0.3\textwidth, angle=-90]
{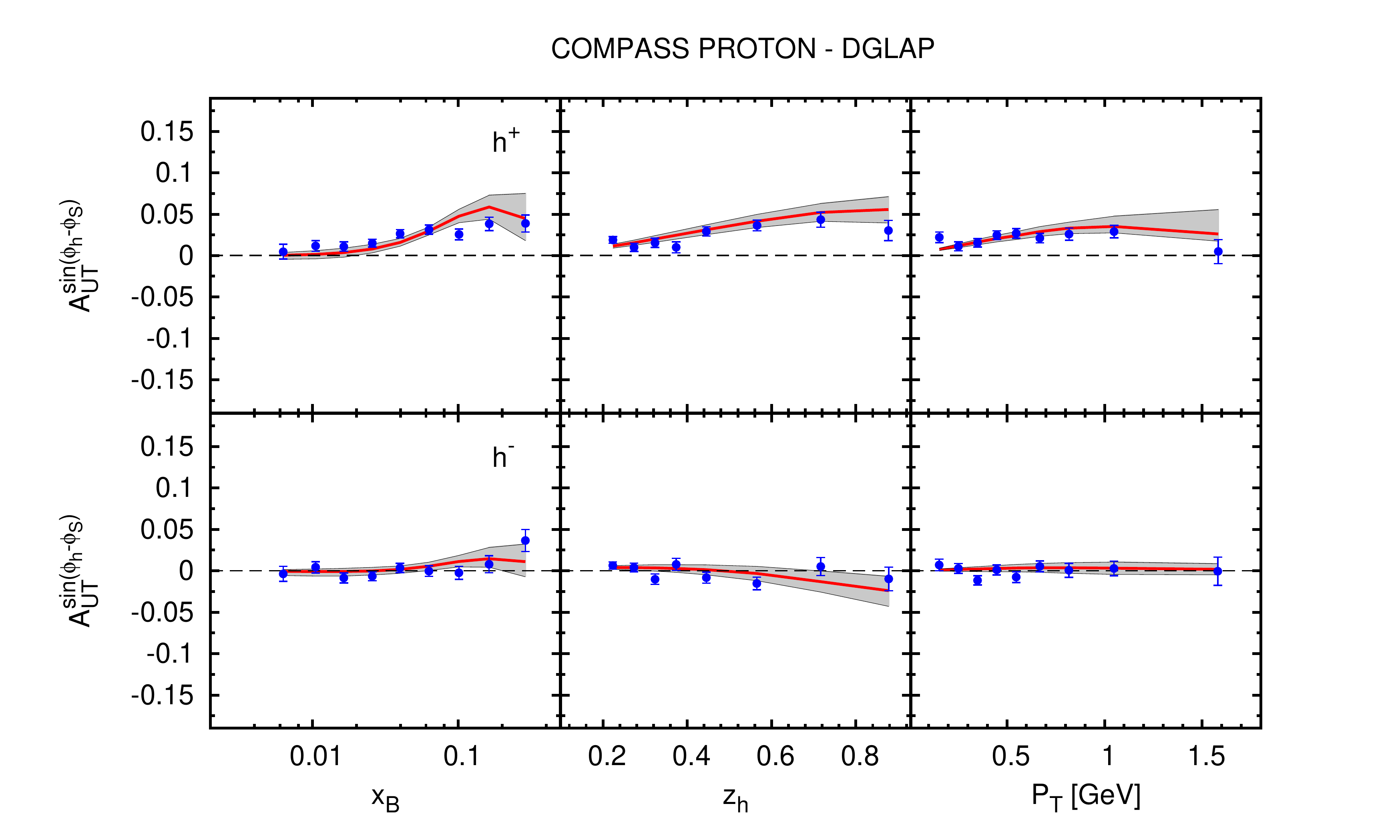}
\caption{\label{fig:compass-p-pi}
The results obtained from the TMD-evolution fit (left panel) and from the  
DGLAP-evolution fit (right panel) of the SIDIS $A_{UT}^{\sin{(\phi_h-\phi_S)}}$ 
Sivers asymmetries (red, solid lines) are compared with the COMPASS-p 
experimental data~\cite{Bradamante:2011xu} for charged hadron production. 
The shaded area corresponds to the statistical uncertainty of the parameters, 
see Appendix~A of Ref.~\cite{Anselmino:2008sga} for further details.
}
\end{center}
\end{figure}
%

In Figs.~\ref{fig:hermes-pi} we show, as an illustration of their qualities, 
our best fits (solid red lines) of the HERMES experimental data~\cite{:2009ti} 
on the Sivers asymmetries for pion production. Those on the left panels  
are obtained adopting the new TMD evolution, while those on the right use the
simplified DGLAP evolution. Similar results are shown, for the recent COMPASS 
data off a proton target~\cite{Bradamante:2011xu} for charged hadron 
production, in Fig.~\ref{fig:compass-p-pi}.    

The shaded area represents the statistical uncertainty of the fit parameters corresponding to a $\Delta \chi^2=20$ ({\it i.e.} to $95.45\%$ confidence 
level for 11 degrees of freedom, see Appendix~A of 
Ref.~\cite{Anselmino:2008sga} for further details). Notice that, in general, 
the error bands corresponding to the TMD-evolution fit are thinner than those corresponding to the DGLAP fit: this is caused by the fact that the TMD 
evolution implies a ratio Sivers/PDF which becomes smaller with growing $Q^2$, 
as shown in Fig.~\ref{fig:siv-confronto}, constraining the free parameters 
much more tightly than in the DGLAP-evolution fit, where the Sivers/PDF 
ratio remains roughly constant as $Q^2$ raises from low to large values.

\begin{figure}[t]
\begin{center}
\includegraphics[width=0.5\textwidth, angle=-90]{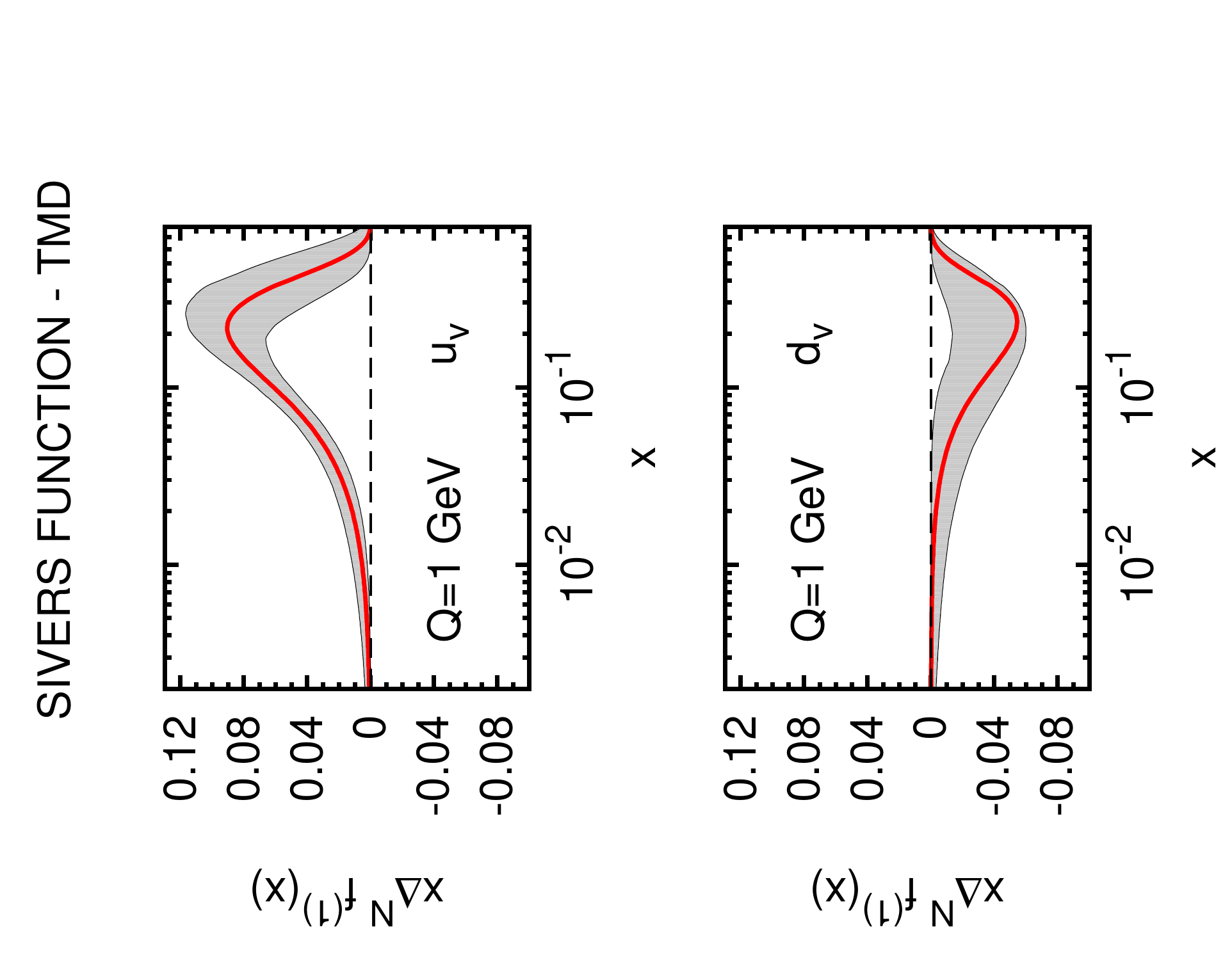}
\hspace*{.3cm}
\includegraphics[width=0.5\textwidth, angle=-90]{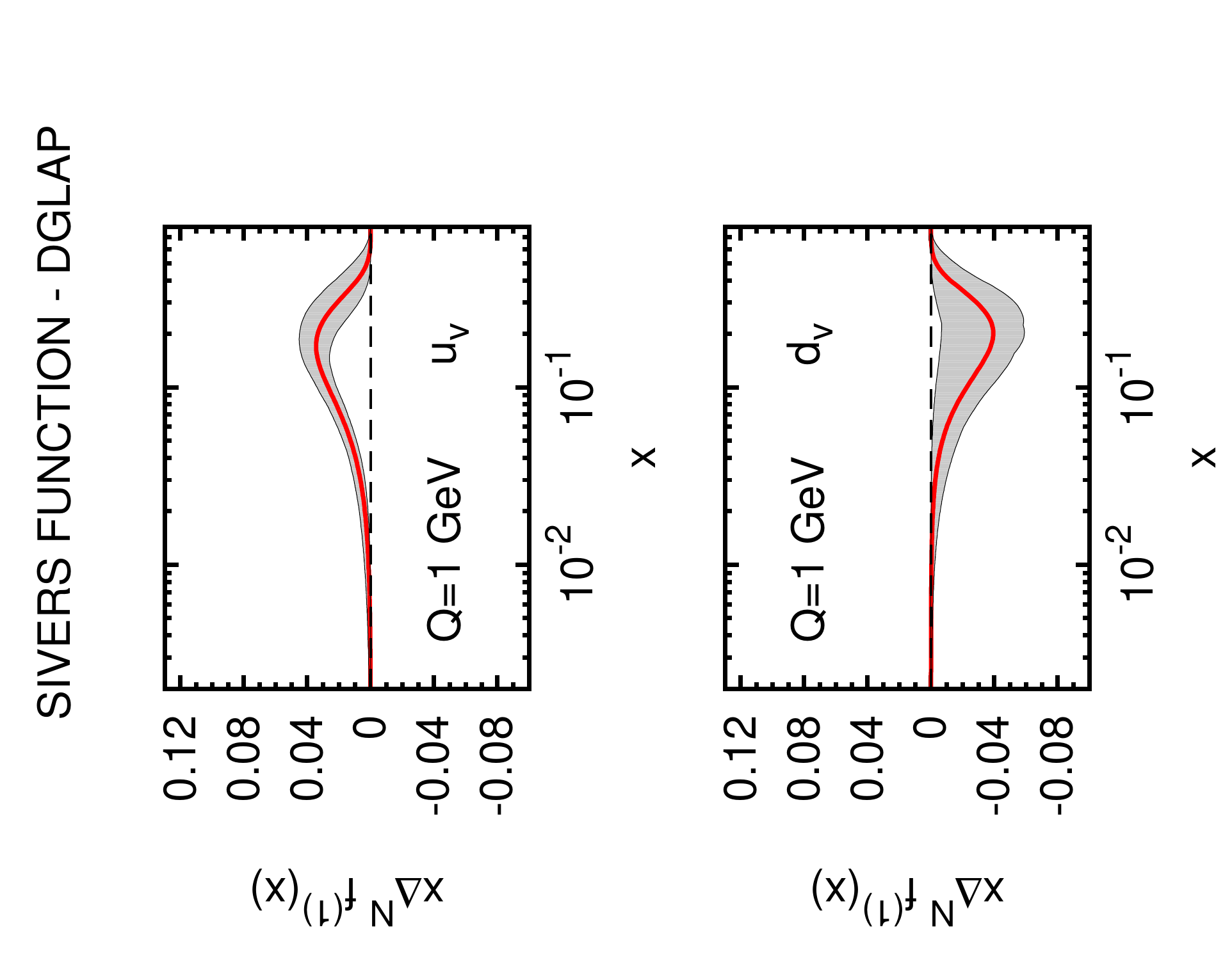}
\caption{\label{fig:siv-tmd-Q0}
The first moment of the valence $u$ and $d$ Sivers functions, evaluated at 
$Q=Q_0$, obtained from our best fits of the $A_{UT}^{\sin{(\phi_h-\phi_S)}}$ 
azimuthal moments as measured by HERMES~\cite{:2009ti} and 
COMPASS~\cite{Bradamante:2011xu,:2008dn} Collaborations. The extraction 
of the Sivers functions on the left side takes into account the TMD-evolution 
(left column of Table~\ref{tab:fitpar_sivers-tmd}), while for those on the 
right side it does not (right column of Table~\ref{tab:fitpar_sivers-tmd}).   
The shaded area corresponds to the statistical uncertainty of the parameters, 
see Appendix~A of Ref.~\cite{Anselmino:2008sga} for further details.
}
\end{center}
\end{figure}

In Fig.~\ref{fig:siv-tmd-Q0} we compare, for illustration purposes, the Sivers function -- actually, its first moment, defined in Ref.~\cite{Anselmino:2008sga} 
-- at the initial scale $Q_0$ for $u$ and $d$ valence quarks, as obtained 
in our best fits with the TMD (left panel) and the DGLAP (right panel) 
evolution, Table~\ref{tab:fitpar_sivers-tmd}. Notice that for this analysis 
we have chosen to separate valence from sea quark contributions, while in 
Ref.~\cite{Anselmino:2008sga} the $u$ and $d$ flavors included all 
contributions.  
%
%
\begin{table}[t]
\caption{Best values of the free parameters, Eq.~(\ref{par_broken}), for 
the Sivers functions of $u$ and $d$ valence quarks, as obtained from our 
TMD-fit, TMD-analytical-fit and DGLAP-fit, at $Q_0 = 1$ GeV. The errors 
reported in this table correspond to the maximum and minimum values of each 
parameter in a restricted parameter space constrained by the condition 
$\Delta \chi^2=20$, corresponding to $95.45\%$ confidence level. 
They correspond to the shaded area in Fig.~\ref{fig:siv-tmd-Q0}.
\label{tab:fitpar_sivers-tmd}}
\vspace*{6pt}
\begin{ruledtabular}
\begin{tabular}{rcl}
\\
 TMD Evolution (exact)   & TMD Evolution (analytical) & DGLAP Evolution\\
\hline\\
$N_{u_v} = 0.77 ^{+ 0.23}_{-0.19}$& $N_{u_v} = 0.75 ^{+ 0.25}_{-0.21} $&$N_{u_v} = 0.45^{+ 0.25}_{-0.17}$\\
\\
$N_{d_v} = -1.00 ^{+ 0.75}_{-0.00}$&$N_{d_v} = -1.00 ^{+ 0.82}_{-0.00}$&$N_{d_v} = -1.00 ^{+ 0.85}_{-0.00}$\\
\\
$\alpha _{u_v} = 0.68 ^{+ 0.57}_{-0.40}$&$\alpha _{u_v} = 0.82 ^{+ 0.51}_{-0.48}$&$\alpha _{u_v} = 1.08 ^{+ 0.68}_{-0.62}$\\
\\
$\alpha_{d_v} = 1.11 ^{+ 1.39}_{-0.91}$&$\alpha_{d_v} = 1.36 ^{+ 1.24}_{-1.00}$&$\alpha_{d_v} = 1.7 ^{+ 1.15}_{-0.91}$\\
\\
$\beta \;\;= 3.1 ^{+ 4.7}_{-2.6}$&$\beta \;\;= 4.0 ^{+ 4.5}_{-2.8}$&$\beta \;\;= 6.9 ^{+ 6.4}_{-4.1}$\\
\\
$M_1^2 = 0.40 ^{+ 1.5}_{-0.23}$ GeV$^2$ &$M_1^2 = 0.34 ^{+ 1.36}_{-0.19}$ GeV$^2$ 
& $M_1^2 = 0.19 ^{+ 0.77}_{-0.10}$ GeV$^2$
\end{tabular}
\end{ruledtabular}
\end{table}

This result deserves some comments. The comparison shows that the extracted 
$u$ and $d$ valence contributions, at the initial scale $Q_0 = 1$ GeV, are 
definitely larger for the TMD evolution fit. This reflects the TMD evolution 
property, according to which the Sivers functions are strongly suppressed with 
increasing $Q^2$, which is not the case for the almost static collinear DGLAP 
evolution. Thus, in order to fit the same data at $Q^2$ bins ranging from 1.3
to 20.5 GeV$^2$, the TMD evolving Sivers functions must start from higher 
values at $Q_0 = 1$ GeV. The Sivers distributions previously extracted, with 
the DGLAP evolution, in Refs.~\cite{Anselmino:2008sga,Anselmino:2011gs} were 
given at $Q^2 = 2.4$~GeV$^2$; one should notice that if we TMD evolve the 
Sivers distributions on the left side of Fig.~\ref{fig:siv-tmd-Q0} up to 
$Q^2 = 2.4$~GeV$^2$ we would obtain a result very close to that of
Refs.~\cite{Anselmino:2008sga,Anselmino:2011gs} (and to that of the right
side of Fig.~\ref{fig:siv-tmd-Q0}).
  
\section{\label{conc} Conclusions and further remarks}

We have addressed the issue of testing whether or not the recently proposed 
$Q^2$ evolution of the TMDs (TMD-evolution) can already be observed in the  
available SIDIS data on the Sivers asymmetry. It is a first crucial step 
towards the implementation, based on the TMD-evolution equations of 
Refs.~\cite{Collins:2011book,Aybat:2011zv,Aybat:2011ge}, of a consistent 
QCD framework in which to study the TMDs and their full $Q^2$ dependence. 
That would put the study of TMDs -- and the related reconstruction of the 
3-dimensional parton momentum structure of the nucleons -- on a firm basis, 
comparable to that used for the integrated PDFs.      

Previous extractions of the Sivers functions from SIDIS data included 
some simplified treatment of the $Q^2$ evolution, which essentially 
amounted to consider the evolution of the collinear and factorized part 
of the distribution and fragmentation functions (DGLAP-evolution). 
It induced modest effects, because of the slow $Q^2$ evolution and of the 
limited $Q^2$ range spanned by the available data. The situation has 
recently much progressed, for two reasons: the new 
TMD-evolution~\cite{Aybat:2011zv,Aybat:2011ge} shows a strong 
variation with $Q^2$ of the functional form of the unpolarized and Sivers 
TMDs, as functions of the intrinsic momentum $\kt$; in addition, some new 
COMPASS results give access to Sivers asymmetries at larger $Q^2$ values.         

It appears then possible to test the new TMD-evolution. In order to do so 
one has to implement the full machinery of the TMD-evolution equations in a 
viable phenomenological scheme. We have done so following 
Ref.~\cite{Aybat:2011ge} and the simplified version of the TMD-evolution
given in Eqs.~(\ref{Ftev})-(\ref{RQQ0}). We have used them in our 
previous procedure adopted for the extraction of the Sivers 
functions~\cite{Anselmino:2008sga,Anselmino:2011gs,Anselmino:2011ch}, with
the same input parameters; moreover, we have considered also the updated 
HERMES~\cite{:2009ti} and the new COMPASS~\cite{Bradamante:2011xu} data.     

A definite statement resulting from our analysis is that the best fit 
of all SIDIS data on the Sivers asymmetry using TMD-evolution, when compared 
with the same analysis performed with the simplified DGLAP-evolution, exhibits 
a smaller value of the total $\chi^2$, as shown in Table~\ref{tab:chi-sq}.
Not only, but when analyzing the partial contributions to the total $\chi^2$ 
value of the single subsets of data, one realizes that such a smaller value 
mostly originates from the large $Q^2$ COMPASS data, which are greatly 
affected by the TMD evolution. We consider this as an indication in favor 
of the TMD evolution.  

A more comprehensive study of the TMD evolution and its phenomenological 
implications is now necessary. Both the general scheme and its application 
to physical processes need improvements. The recovery of the usual collinear 
DGLAP evolution equations, after integration of the TMD evolution results
over the intrinsic momenta, has to be understood. Consider, as an example, 
the simple expression of the evolution of the unpolarized TMD PDF, as given 
in Eq.~(\ref{unp-gauss-evol}). Such an evolution describes how the TMD 
dependence on $\kt$ changes with $Q^2$, but does not induce any change in 
the $x$ dependence, which, at this order, remains fixed and factorized. The 
question whether or not one can recover the usual DGLAP evolution, which 
changes the $x$ dependence, for the integrated PDFs arises naturally at 
this point. A naive integration of Eq.~(\ref{unp-gauss-evol}) on $\bkt$, 
over the full integration range, would give $f_{q/p}(x,Q) =
f_{q/p}(x,Q_0) \, R(Q,Q_0)$ which is not the correct PDF evolution. However,
the $\bkt$ integration should have upper limits which depend on $x$ and $Q^2$,
and the full TMD evolution is more complicated than the simplified version 
used here, as explained at the beginning of Section~\ref{Intro-sub}. 

We have made a safe phenomenological use of the TMD evolution equations; 
it is true that they induce a strong change in the 
$\kt$ dependence of the unpolarized and Sivers TMDs, leaving unchanged the 
$x$ dependent shape, thus neglecting the collinear DGLAP evolution, but this 
should not be a problem. Infact, as we have shown explicitly in 
Fig.~\ref{fig:unp-confronto} (dashed curve), the collinear DGLAP evolution 
is negligible in the $Q^2$ region considered; Fig.~\ref{fig:unp-confronto} 
is drawn for $x = 0.1$, but a similar conclusion holds for all $x$ values 
involved in the SIDIS data used in the paper. Moreover, the extra factors 
$R(Q,Q_0)$ arising in the TMD evolution, cancel out, as already explained, 
in the expression of the Sivers asymmetries.

A fresh analysis of TMD dependent data, both in polarized and unpolarized, 
SIDIS and Drell-Yan processes, has to be carefully performed including 
TMD evolution from the beginning in an unbiased way. Most importantly so, 
should predictions for future high energy experiments, like the planned EIC/ENC 
colliders, be considered or re-considered.       

\section{Acknowledgements}

We thank Umberto D'Alesio, Francesco Murgia and Alexei Prokudin for 
interesting and fruitful discussions.
We acknowledge support of the European Community under the FP7 
``Capacities - Research Infrastructures'' program (HadronPhysics3, 
Grant Agreement 283286).
We acknowledge partial support by MIUR under Cofinanziamento PRIN 2008.

%

\end{document}